# Learning to measure resistance noise demystifies the ubiquitous 1/$f$ excess noise.

**J. I. Izpura**

**Abstract**

To study resistance noise ($\Delta R$) by a spectrum analyzer we must convert this noise into a noise voltage ($\Delta V$) at the reach of such generalized voltmeter. Whenever a current $I_{conv}$ is set in a resistor to convert its resistance noise into noise voltage by Ohm's Law: $\Delta V = \Delta R \times I_{conv}$, the converted noise thus obtained does not track $\Delta R_{TE}$ (its resistance noise in Thermal Equilibrium, TE) but $\Delta R$, that is: a resistance noise out of TE due to $I_{conv}$ itself. Thus, backgating noises in the channel of resistors (i. e. Field-Induced Resistance Noise, FIRN) found by this method always are noises out of TE. The way the Lorentzian $\Delta R_{TE}$ of a resistor is converted by $I_{conv}$ into nine decades of resistance noise $\Delta R$ with 1/$f$ spectrum is the lesson we give on this unexpected spectral change that we could express as: "To measure is to disturb, particularly in resistance noise measurements".

# I- Introduction

A resistor in Thermal Equilibrium (TE) at temperature $T$ shows a fluctuating voltage between its terminals with spectral density $S_V(f)=4kTR$ V²/Hz, where $k$ is the Boltzmann constant and $R$ is its resistance, as reported by Johnson [1] and explained by Nyquist [2] in 1928. This *voltage we measure* called Johnson noise is the effect of $S_I(f)=4kT/R$ A²/Hz, which is the *theoretical* density of Nyquist noise that we do not measure, but only infer from $S_V(f)$ [3]. This comes from the fast conversion of $S_I(f)$ into $S_V(f)$ (integration) carried out by $C$, the non-null capacity between terminals of any resistor. Biasing a resistor by a dc current $I_{conv}$, two noise voltages appear between its terminals: **i)** the same density $S_V(f)$ it showed in TE when heating effects of $I_{conv}$ are negligible (Johnson floor) and **ii)** a new noise voltage emerging over this floor at low frequencies, thus an "*excess noise*" that *only is observed while $I_{conv}$ exists* (see Fig. 1).

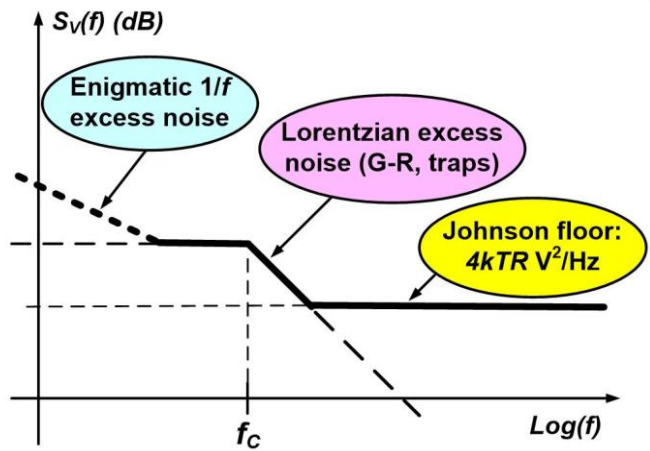

**Figure 1.** Spectral density of noise voltage at low frequencies that can be found in a resistor of resistance $R$ biased by a current $I_{conv}$ whose heating effects are negligible.

Although the presence of the Johnson floor when the resistor roughly keeps the same $T$ it had in TE is not surprising from [1, 2], the lack of the noise voltage due to the shot noise $S_I(f)=2qI_{conv}$ A²/Hz assigned to $I_{conv}$ today is a "surprise" that vanishes from the baseless assignment of shot noise to *conduction currents* [3] following a fluctuation-based noise model [4, 5] that we will use hereafter for devices out of TE, where the dissipation-based noise model currently used hardly applies. Since for an $I_{conv}$ not coming from displacement currents $S_I(f)$ would be null, this rules out shot noise as a possible source of the excess noise of Fig. 1.

As it is widely accepted today, excess noise reveals small fluctuations of resistance (resistance noise $\Delta R$) that exist in resistors regardless their technology [6, 7]. Thus, excess noise is a general feature of the two-terminal device (2TD) we call resistor, which always offers between its terminals some resistance noise $\Delta R$ around its *mean* resistance value $R$. This resistance noise that given by its root mean square ($\Delta R$ in $\Omega_{rms}$) divided by $R$ uses to fall in the 0.1 ppm range or below (i. e. $\Delta R/R \leq 10^{-7}$), would be revealing an spectrum of resistance noise $\Delta R$ in the resistor because its spectral density (in V²/Hz) is proportional to $(I_{conv})^2$, as it should be from Ohm's Law. Taking mean square values, we would expect:

$$\Delta V = \Delta R \times I_{conv} \Rightarrow \langle(\Delta V)^2\rangle = (I_{conv})^2 \times \langle(\Delta R)^2\rangle \qquad (1)$$

Thus, no matter if excess noise shows a 1/*f* spectrum (*f* is frequency) or not, it would come from resistance noise Δ*R* that $I_{conv}$ *reveals* as noise voltage Δ*V*. This shows the method we use to study Δ*R* in resistors: by biasing them with a dc current $I_{conv}$ high enough so as to get afloat some excess noise Δ*V* over the Johnson floor of the resistor. The spectrum of Δ*V* thus found will mirror that of resistance noise Δ*R* existing in the resistor with a conversion gain $G_{conv}=(I_{conv})^2$, but the fact that most people ignore is that the resistance noise Δ*R* existing while $I_{conv}$ puts the resistor out of TE is not the noise $\Delta R_{TE}$ that existed previously in TE. This is the key notion to understand how the ubiquitous 1/*f* excess noise appears in solid-state devices. We gave it in a quite specialized paper [6] that now we are putting at the reach of a wider audience by the set of new ideas on resistance noise and its measurement that we show in this new paper.

To follow this work, readers should know the *field effect* that governs the so called Field Effect Transistor (FET) by which a Bordering Space-Charge Region (BSCR) modulates the cross section of its conductive channel. Changes in Φ, the built-in potential of this BSCR, lead to changes in its width that vary the thickness and the resistance $R_{ch}$ of the adjacent channel. Hence, if an applied voltage on the input capacity $C_{GS}$ of a FET modulates its $R_{ch}$ by varying the cross section of its channel, *the own thermal voltage of $C_{GS}$ will produce a resistance noise $\Delta R_{ch}$* when its gate is left floating. From this notion and substituting some mathematics by figures showing the relevant effects, we will show that the 1/*f* excess noise of resistors is due to resistance noise induced in their conductive channels by field effect from thermal noise born in BSCRs. Because any channel has BSCRs due to interfaces and surfaces, as well as embedded SCRs around dislocations and defects, the Field-Induced Resistance Noise (FIRN) of [6] that we will clarify here becomes a ubiquitous phenomenon whose genesis and proper measurement is worth knowing.

To understand the way FIRN interacts with its measurement procedure to give the puzzling 1/*f* excess noise, Physics and Engineering notions are needed. Although physical reasons for the ubiquity of FIRN have been given, its striking 1/*f* spectrum covering many decades of frequency requires the proper handling of those two type of notions and, likely, an extra effort to remove prejudices and beliefs preventing the proper judgement of new ideas that apparently conflict with well-known ones in use. This last sentence refers to misconceptions that can be acquired from previous works, often taken as scientific dogmas that further works should follow. In our case this has to do with [7], whose title is: "*1/f noise is not surface effect*" and we have just said that surfaces are sources of resistance noise. Apparently, our proposal and [7] would conflict, but no conflict exists when we take with care the two phrases written after such a title: "*1/f noise is inversely proportional to the total number of mobile charge carriers in homogeneous samples. This excludes surface effects as the main source of 1/f noise*" [7].

Given the relevance of [7] in 1/*f* excess noise, our proposal of surfaces that cause or induce this noise could be rejected on the basis of its heretic departure from this well-known work. However, this departure only is apparent and since FETs already were used when [7] was published, let us use their foundations to educate scientists on the origin and measurement of FIRN. From the modulation of the channel of a junction FET (JFET) by a gate on its *surface* we can realize that a random modulation from such a surface electrode (or from similar ones existing in the device) will produce a resistance noise in its channel. Thus, the notion that the title of [7] could suggest about surfaces having nothing to do with excess noise must be replaced by the likely meaning of a subsequent reasoning built on its "premise": "*1/f noise is inversely proportional to the total number of mobile charge carriers in homogeneous samples.*" and its "concluding sentence": "*This excludes surface effects as the main source of 1/f noise*".

This premise stating that excess noise *senses the size of the device* where it is measured by its total number N of carriers means that small devices with few carriers in their small volume should show higher excess noise than big ones, as it happens when devices are shrunk to increase the device density of integrated circuits. As we will show, FIRN agrees very well with this feature of small devices. Thus, our apparent conflict with [7] could come from the lack of an explicit second premise to build a syllogism leading to a logical conclusion from such premises. Adding as second premise: "*Surface currents do not involve the total number N of carriers in the device where we measure excess noise*" a logical conclusion could be: "*This excludes surface currents as the main source of 1/f noise*". From this result pointing towards excess noise having to do with currents *in the bulk* region of the device, thus far from its surfaces, most people tend to believe that they are measuring excess noise in an unlimited bulk of material devoid of borders, thus overlooking *the device* where they actually are measuring this noise voltage.

The employment of the word "samples" instead of "devices" uses to be a symptom of this oversight, but from [3-6] it should be clear that the noise we measure is determined by the device we use rather than by its inner material. This inner material can be "none" as we showed in [8] for the device made by two metallic plates in TE at temperature *T*, kept at a distance *d* in vacuum. Going back to [7] let us consider its empirical formula that summarized excess noise with 1/*f* spectrum found in a broad set of resistors up to the year 1969. This formula links the total number of carriers (N) involved in the electrical conduction of each resistor with $S_R(f)=\langle(\Delta R_{ch})^2\rangle/R_{ch}^2$, which is the spectral power density of normalized fluctuations of resistance. This $S_R(f)$ is deduced from the 1/*f* excess noise of the resistor measured while a current like $I_{conv}$ exists in this device. We refer to Hooge's formula, which is:

$$S_R(f) = \frac{\langle(\Delta V)^2\rangle}{V_{DS}^2} = \frac{I_{conv}^2 \times \langle(\Delta R_{ch})^2\rangle}{I_{conv}^2 \times R_{ch}^2} = \frac{\langle(\Delta R_{ch})^2\rangle}{R_{ch}^2} \approx \alpha_H \times \frac{1}{f \times N} \qquad (2)$$

Given the high number of different $I_{conv}$ values considered in [7], an added merit of Eq. (2) is the normalized form it has to summarize the average trend of a wide set of scattered measurements. Because every resistor has a conductive channel between its two terminals, we have used $R_{ch}$ for its average resistance and $\Delta R_{ch}$ for its resistance noise. These subscripts do not appear in [7], where no explicit mention appears about the channeled structure of any resistor. However, this notion is implicit in the volume of material we must consider to calculate the total number of carriers N involved in the electrical conduction of this device. To the best of our knowledge, all the data leading to Eq. (2) came from excess noise found while a current like $I_{conv}$ was set in the channel of each resistor under test. Thus, Eq. (2) could be revealing a *general trend of the excess noise created in the channel of each resistor by this technique* that uses $I_{conv}$ to convert resistance noise $\Delta R_{ch}$ into noise voltage $\Delta V$ (its excess noise). This was the radical message of [6] for experts in "1/*f* noise" that needs to be put at the reach of a wider set of scientists working in the electrical noise field.

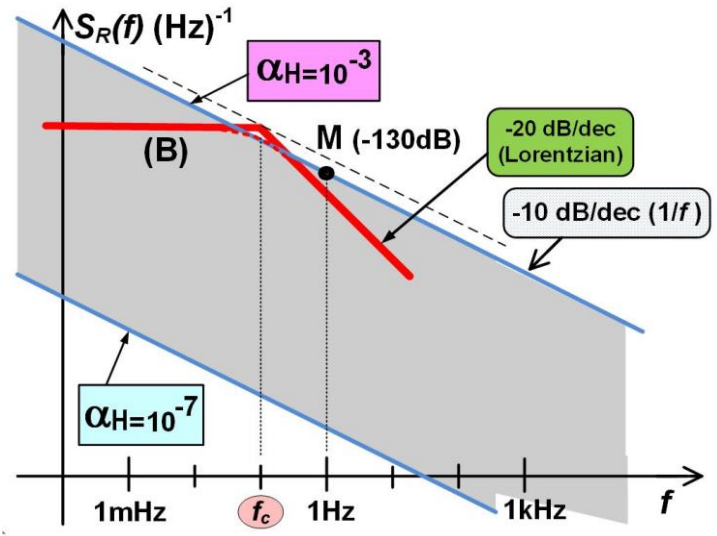

**Figure 2.** Region obtained from Hooge's formula for the spectral power density $S_R(f)$ deduced from excess noise measurements taken in resistors biased by a current $I_{conv}$. Graph (B) will be described at a later time.

Following Eq. (2) the 1/*f* excess noise we should find in a resistor (see the "*f*" in its denominator) should be inversely proportional to its total carrier number N through the dimensionless Hooge's "constant" $\alpha_H \approx 2 \times 10^{-3}$ [7] that a more recent view on the subject sets in the range $10^{-7} < \alpha_H < 10^{-3}$ [9]. Thus, Eq. (2) with this range of $\alpha_H$ values should account for the spectral power density $S_R(f)$ (Hz$^{-1}$) of normalized fluctuations of resistance deduced from measurements of 1/*f* excess noise in *resistors biased by a conduction current $I_{conv}$*. This means that the $S_R(f)$ deduced from the excess noise of resistors should show a 1/*f* spectrum falling in the shadowed band of Fig. 2. This range of values for $\alpha_H$ likely comes from the broad set of $I_{conv}$ values used by different authors and from the different materials, geometries and doping levels of their *devices* under test.

Using Eq. (2), point M giving $S_R(1Hz)$=-130 dB at 1Hz appears for a resistor with N=$10^{10}$ carriers in its channel made from a rather noisy technology leading to $\alpha_H=10^{-3}$. Let us imagine that this channel is a 1μm thick layer of semiconductor doped with $10^{17}$ carriers/cm$^3$ and cut in the form of a strip of length L=1mm and width W=100µm. A resistor of the same $R_{ch}$ but four times noisier would result by halving both L and W, see Eq. (3) below. From Eq. (2) this action would give rise to a new device with N/4 carriers, thus four times noisier. This prediction of Eq. (2) agrees well with the behavior of the FIRN we will consider below. This feature together with the ability of FIRN to give many decades of excess noise with $1/f$ spectrum strongly suggest that FIRN gives the elusive explanation for the $1/f$ excess noise of resistors.

Because our target audience are scientists involved in noise measurements rather than experts in $1/f$ excess noise, a good book to follow this work is [10] (and references therein) due to its broad coverage of electrical noise in solid state devices. Note the relevance we give to the *device* where we measure this noise voltage because it determines the result we obtain (spectrum included) as we will show. This is a result that should help to reduce the dogmatism often found in research, illustrated by this sentence of an anonymous referee on one of our works: "*To my opinion, the 1/f noise problem is of fundamental nature and the circuit approach does not suit for its solution*". By contrast, we believe and we will demonstrate below that if we are measuring $1/f$ noise *voltage* in a *resistor* driven by a front-end electronics, both the "$1/f$ problem" as well as its solution should come from the way this device interacts with its electronics and with its thermal bath, but not from the beliefs of each one on the subject.

## II- The physical structure of resistors and its electrical resistance

Fig. 3 shows the basic model of a resistor of resistance $R_{ch}$ with two parallel terminals (metallic plates) of area $A_{2D}$=W×H separated by a distance L (the length of its conductive channel). Working at low frequency (typically $f$<1MHz) to neglect the unavoidable capacitive effects between terminals of this conductive channel [6] and for σ=1/ρ being the conductivity of its inner material, the resistance $R_{ch}$ is:

$$R_{ch} = \rho \times \frac{L}{W \times H} = \frac{1}{\sigma} \times \frac{L}{A_{2D}} \; \Omega \qquad (3)$$

**Figure 3.** Prismatic channel of conductive material of a resistor ready to be driven by a dc current $I_{conv}$ to convert its resistance $R_{ch}$ and its fluctuations $\Delta R_{ch}$ into *dc* and *ac* voltages respectively.

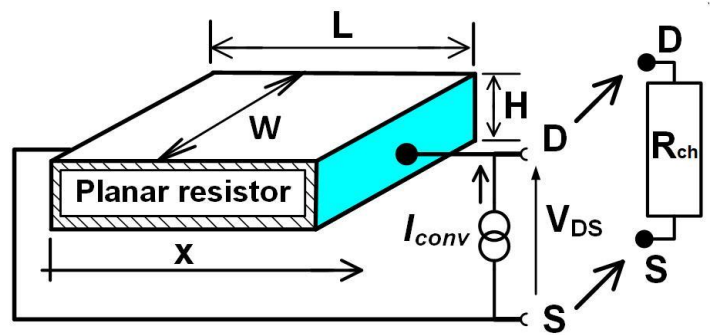

Eq. (3) is found when the electric field within the resistor is orthogonal to its two equipotential terminals. This 1-D model works well for most resistors and particularly for planar ones, where conductive layers with this prismatic form are often found. However, the volume $V_{ch}$ of conductive material giving rise to the $R_{ch}$ found between terminals is not exactly equal to the volume $V_{3D}=W\times H\times L$ of the "slab" of material shown in Fig. 3. This is so because the surfaces limiting the material use to have associated BSCRs taking the form of thin "skins" of non-conductive material under them. N-type GaAs use to have a sheet of negative charge fixed in their surface states that is counterbalanced by the positive one of a depleted "skin" of thickness $t_{wrap}\approx 0{,}1\mu m$ underneath. Thus, this skin is a thin layer of fixed charge due to unscreened, ionized donor atoms that do not take part in the electrical conduction between terminals [11].

In this way the volume of the conductive channel of GaAs ($V_{ch}$) is slightly lower than the volume $V_{3D}$ of GaAs. Since this *surface effect* that reduces the cross section of the channel does not change its length, the conductivity of the inner GaAs inferred by Eq. (3) from a measurement of $R_{ch}$ will give a slightly lower conductivity than the actual one because the cross section $A_{2D}$ of the slab of material appearing in Eq. (3) is slightly higher than the actual cross section of the channel. For W and H falling in the mm range, the relative error in σ due to this non conducting skin of thickness $t_{wrap}\approx 10^{-4}$ mm would be in the tens of ppm (parts per million). Thus, the σ obtained by neglecting $t_{wrap}$ is accurate enough for most purposes, but for epitaxial layers with H≈1μm the relative error reaches the 1%-10% range. This shows why the electrical structure of the device under test is so important, particularly to deal with ∆R/R noises under the ppm range. From Eq. (3) and Fig. 3, small relative changes ∆H/H in the thickness of the channel should produce small relative changes $\Delta R_{ch}/R_{ch}$ in its resistance. Differentiating Eq. (3) respect to the thickness H and rearranging terms we obtain:

$$R_{ch} = \rho \frac{L}{W\times H} \Rightarrow \frac{\Delta R_{ch}}{R_{ch}} = -\frac{\Delta H}{H} \Rightarrow \Delta R_{ch} = \left|\frac{\Delta H}{H}\right| \times R_{ch} \qquad (4)$$

Although its sign is irrelevant, the *linear relationship* set by Eq. (4) between fluctuations ∆H and $\Delta R_{ch}$ when $\Delta R_{ch}/R_{ch}$ is in the ppm range is important. Such being the case, a series of tiny changes of thickness ∆H is linearly converted into a series of small changes $\Delta R_{ch}$ of $R_{ch}$. Thus, a Lorentzian spectrum of $\Delta R_{ch}$ found in a resistor means that its channel *or a region of it*, is undergoing a fluctuation of its thickness with Lorentzian spectrum. Note that a noise $\Delta R_{ch}$ measured between the two terminals of a long conductive channel *does not tell us "where" within the channel such noise is being generated*. Sections to come will exploit this fact and because planar resistors are not "isolated channels floating in space" as Fig. 3 could suggest, let us consider this fact it in the next Section.

### III- Field induced resistance noise (FIRN): the noise of the small layouts

Due to our acquaintance with GaAs devices and to the opportunity to take empirical data from [11] that inspired [6], we will use GaAs resistors hereafter although their geometry will recall devices of Si planar technology where semi-insulating (SI) substrates like those of GaAs technology are not available. Thus, let us consider the resistor of Fig. 4 made from a p$^+$-GaAs substrate giving rigidity to an n-GaAs layer on top that shares a bordering surface with the p$^+$-GaAs of the substrate and another bordering surface with the air on top (its naked surface).

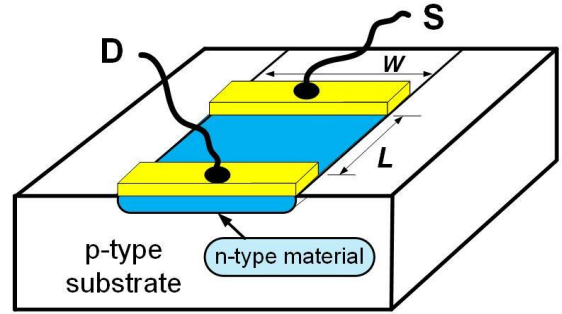

**Figure 4.** Physical structure of a resistor made from a layer of n-type GaAs onto a p$^+$-type GaAs substrate (see the text).

Although the p$^+$-GaAs that is under the thin n-GaAs layer only is used for rigidity purposes, *it is there* together with its *proximity effect* on the channel called *backgating* that is easier to calculate than the *weaker but non-null backgating* from a SI-GaAs substrate like those of [11]. Fig. 5 shows this n-GaAs channel resting against a p$^+$-GaAs layer in a vertical view of these two layers side by side to show that every GaAs resistor made from this "planar technology" is a long FET whose gate is the p$^+$-GaAs substrate and whose channel is the n-GaAs layer on top. If our resistor was in an integrated circuit, its substrate would be biased by the lowest possible voltage to provide "dc isolation" by keeping reverse-biased the p$^+$-n junction under its channel. Shorting the S terminal and the p$^+$-GaAs substrate we could "silence" this backgating effect, but this resistor becoming a diode is not acceptable. Because we only need a single resistor we will leave *floating* the p$^+$-GaAs substrate to avoid such diode effect.

Thus, our resistor is a long FET whose channel offers its mean resistance $R_{ch}$ between terminals together with its resistance noise $\Delta R_{ch}$ fluctuating around $R_{ch}$. People working in electrical noise use to know that a FET with its gate left floating is prone to collect electromagnetic noise by its gate wire acting as an antenna. Small voltages collected in this way would appear on the input capacity $C_{GS}$ of the long FET that our resistor is, thus modulating the thickness of its channel around $H_{eff}$ (its mean value). This $H_{eff}$ will be lower than H to give room for the width $t_{back}$ of the one-sided BSCR of this p$^+$-n junction that mainly lies in the less doped n-GaAs side. It is worth noting that the Faraday cage used in noise setups and that would suppress the antenna effect of a floating gate, is useless to *prevent the FIRN* due to the thermal noise of $C_{GS}$ because its $kT/C_{GS}$ noise [10] *already is there* due to the thermal bath of our device [4, 5].

To handle real data of GaAs devices let us consider the photo-resistors made from a 1 µm thick n-GaAs layer that we described and measured in [11]. Although these devices had SI-GaAs substrates instead p$^+$-GaAs ones, this did not avoid the BSCR between the n-GaAs and the substrate nor its backgating effect on $R_{ch}$. The action of the two BSCR cladding the n-GaAs layer of these devices was unambiguously shown in [11] to give a cogent explanation for the high, but slow gain of these photo-detectors that was different from their photo-conductive gain at high illumination levels, hence the title of [11]. Thinking of the noise that limited the signal/noise ratio of this high but slow gain, we concluded that it was resistance noise thermally generated by the same mechanism giving rise to this gain: fluctuations of their channel thickness ($\Delta H_{eff}$). This led us to study the thermal noise of their (substrate-channel) and (gate-channel) junctions under open circuit conditions and this is why we propose Fig. 5 to study the thermal noises of the JFET that the "simple" resistor of Fig. 4 is.

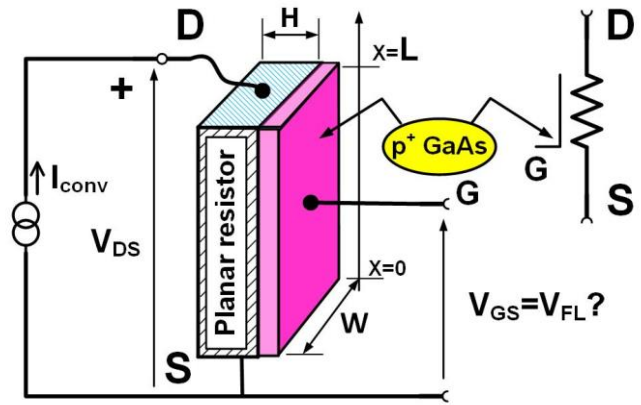

**Figure 5.** Electrical structure of the resistor made from a layer of n-GaAs onto a GaAs substrate (a p$^+$-GaAs one, see the text).

Taking the n-GaAs/p$^+$-GaAs diode, its input admittance for thermal noise is due to its capacity $C_{GS}$ shunted by $R_{GS}$ (its differential resistance) [4, 5]. Shorting the substrate to the S terminal we could "silence" this backgating effect, but the diode thus formed is not acceptable and not all sources of resistance noise can be silenced in this way. An example is the "topgating" effect due to the upper SCR of thickness $t_{wrap}$ associated to the naked GaAs-air interface of Fig. 4. Although we will take into account $t_{wrap}$ for a closer dc modelling of our devices, we will not consider the FIRN due to this upper BSCR that, being able to store electrostatic energy, will show small fluctuations of $t_{wrap}$ to accomplish thermal equipartition [12]. With regard $\Delta R_{ch}$, the FIRN generated in the device of Fig. 5 by its floating p$^+$-GaAs substrate will be high because using p$^+$-GaAs substrates for mechanical rigidity is a "noisy technology" leading to a high $\alpha_H$ for the FIRN with $1/f$ spectrum that our device of Fig. 5 will show when biased by $I_{conv}$. Let us begin our reasoning from $C_{GS}$, the capacity of the p$^+$-n diode of Fig. 5 and its dynamical resistance $R_{GS}$ (both in TE) whose parallel connection forms the admittance that defines the Lorentzian spectrum of noise voltage of this device [4, 5]. From $f_c=1/(2\pi R_{GS}C_{GS})$ the spectrum of noise voltage in $C_{GS}$ will be this one:

$$S_V(f) = \frac{S_{V0}}{1+\left(f/f_c\right)^2} = \frac{4kTR_{GS}}{1+\left(f/f_c\right)^2} \; V^2/Hz \qquad (5)$$

Integrating Eq. (5) from $f \to 0$ to $f \to \infty$ or using its equivalent noise bandwidth $BW_N=(\pi/2)f_c$, we obtain the mean square noise voltage in $C_{GS}$ known as the $kT/C$ noise (in $V^2$) of this capacitor in TE, which does not depend on its shunting $R_{GS}$ because it appears by applying thermal equipartition to the degree of freedom that $C_{GS}$ represents [10]. Therefore:

$$\langle (v_{GS})^2 \rangle = \int_0^\infty \frac{4kTR_{GS}}{1+\left(f/f_c\right)^2} df = 4kTR_{GS} \times \frac{\pi f_c}{2} = \frac{kT}{C_{GS}} \; V^2 \qquad (6)$$

With regard the cut-off frequency $f_c$ of the Lorentzian spectrum of Eq. 5 we will say that because the energy barrier ($q\Phi \approx 1,3eV$) of a $p^+$-n GaAs junction diode roughly is 300meV higher than that of a Si one, its time constant given by $\tau_{gs}=(R_{GS} \times C_{GS})$ easily will reach the range of seconds and even minutes. Taking only $\tau_{gs} \approx 1,59$ seconds, the cut-off frequency of this thermal noise of $C_{GS}$ giving rise to FIRN in the channel of the long FET of Fig. 5 would be:

$$f_C = \frac{1}{2\pi(R_{GS} \times C_{GS})} \approx 0,1 \; Hz \qquad (7)$$

Thus, the thermal noise voltage between the n-GaAs layer of our resistor and its $p^+$-GaAs floating substrate mostly will be in a narrow band up to $f_c \approx 0.1$Hz. From $C_{GS} \approx 70$pF (that we will justify below for L=1mm and W=0,1 mm) and Eqs. (6) and (7) we obtain: $\langle (v_{gs})^2 \rangle = 59 \times 10^{-12} \; V^2$ at $T=300$K and $R_{GS} \approx 22,7 \times 10^9 \; \Omega$. In summary: the noise voltage between the channel and its floating Gate in the FET of Fig. 5 in TE would be: $V_{gs} \approx 7,7\mu V_{rms}$. This noise voltage modulating the cross section of this resistor will create a Lorentzian spectrum of resistance noise $\Delta R_{ch}$ that we are going to obtain. From the linear way a FET transfers to its channel thickness small signals driving its input, the spectrum of $\Delta R_{ch}$ should mirror the Lorentzian one of noise voltage in $C_{GS}$ given by Eq. (5). To define the channel of our resistor let us take $N_d \approx 10^{17}$ cm$^{-3}$ as its donor concentration in the n-GaAs layer and H=1µm as its thickness following [11]. Because the $p^+$-n GaAs junction, there will be a *one-sided* SCR of width $t_{back}$ lying in the n-GaAs side, that will reduce the thickness of the n-GaAs channel. Using the built-in voltage $\Phi$ of this $p^+$-n junction, the permittivity $\varepsilon$ of the GaAs and the electronic charge $q$, the thickness $t_{back}$ is given by [11]:

$$t_{back} = \sqrt{\frac{2\varepsilon\Phi}{qN_d}} \qquad (8)$$

From $t_{dark}=0,118$µm given by Eq. (8) for a metal-GaAs junction with $\Phi=0,8$V in [11] we have: $t_{back}=0.15$µm for our $p^+$-n GaAs junction of $\Phi \approx 1,3$V. Thus, the thickness of our n-GaAs channel will be: $(H-t_{back})=0,85$µm in TE. Subtracting $t_{wrap} \approx 0,1$µm of depleted GaAs under the naked surface on top of the n-GaAs layer, the effective thickness for electrical conduction of our n-GaAs channel is: $H_{eff}$=0.75 µm, a value that only is 0,03 µm lower than that of the Hall bars of [11] whose resistance was 2,35kΩ for W=100µm and L=700µm. Using Eq. (3), the channel resistance $R_{ch}$=2,35kΩ for H=0,78µm of [11] would rise up to $R_{ch}$=2,44kΩ for $H_{eff}$=0,75µm. Increasing the length up to L=1000µm while keeping W=100µm and $H_{eff}$=0,75µm, the resistance of the 1mm long resistor obtained in this way is: $R_{ch}$=3,5kΩ, a round number we will use hereafter. For L=1000µm and W=100µm the area of its $p^+$-n diode will be: L×W=0,1 mm$^2$. Using the permittivity $\varepsilon$ of GaAs and $t_{back}$=0,15µm, the capacity $C_{GS}$ of the FET of Fig. 5 will be:

$$C_{GS} = \epsilon \times \frac{L \times W}{t_{back}} \approx 70 \ pF \qquad (9)$$

This justifies $C_{GS}$=70pF we used to obtain $<(v_{gs})^2>$=59×10$^{-12}$ V$^2$ at T=300K. From Eq. (8) giving $t_{back}$≈0,15 µm for Φ≈1,3V, (junction barrier qΦ≈1,3eV) and given that reverse voltages increasing this barrier will add to Φ, we can obtain the voltage $V_p$ to pinch-off the channel of our resistor. To have $t_{back}$≈0,9 µm in Eq. (8) (thus $H_{eff}$=0 in the channel) we need Φ≈47V. Therefore, the voltage $V_p$ required to close the channel is: $V_p$=(47-1,3)≈46V. This means that for voltages $V_{DS}$<2V, the FET of Fig. 5 will be biased well in its "ohmic region", thus *behaving as a resistor* where the thickness of its channel at each position x is quite the same.

Given that any voltage v appearing in $C_{GS}$ (thus in the $p^+$-n junction) adds to Φ when it increases its barrier (reverse bias) or it is subtracted from Φ when it reduces such barrier, Eq. (8) becomes Eq. (10) linking the thickness of the SCR of the $p^+$-n diode with its bias voltage v between the $p^+$-Gate (substrate) and the n-GaAs channel. Differentiating Eq. (8) and particularizing for v=0 we have the channel modulation $\Delta t_{back}$ caused by small fluctuations of voltage in $C_{GS}$ ($\Delta v$) like those coming from its own $kT/C_{GS}$ noise. This appears in the left side of Eq. (10).

$$t_{back}(v) = \sqrt{\frac{2\varepsilon(\Phi-v)}{qN_d}} \Rightarrow \Delta t_{back} = \sqrt{\frac{2\varepsilon\Phi}{qN_d}} \times \frac{\Delta V}{2\Phi} \qquad (10)$$

Thus, the relative change of the BSCR thickness $\Delta t_{back}/t_{back}$ is equal to the relative voltage change $\Delta V/(2\Phi)$ where Φ is the built-in voltage of the $p^+$-n diode. Using mean square values we have: $<(\Delta t_{back})^2>/(t_{back})^2 = <(\Delta V)^2>/4\Phi^2$. From Eq. (4) and since $\Delta H_{eff}/H_{eff} = -\Delta t_{back}/H_{eff}$, the relationship between $<(\Delta R)^2>$, the mean square fluctuation caused in the channel by the mean square noise voltage in the $p^+$-n diode (that is: by $kT/C_{GS}$ V$^2$ in $C_{GS}$) will be:

$$\langle(\Delta R_{ch})^2\rangle = \langle\left|\frac{\Delta t_{back}}{H_{eff}}\right|^2\rangle R_{ch}^2 = \frac{\langle(\Delta V)^2\rangle}{4\Phi^2}\left(\frac{t_{back}}{H_{eff}}\right)^2 R_{ch}^2 = \frac{R_{ch}^2}{4\Phi^2}\left(\frac{0.15}{0.75}\right)^2 \frac{kT}{C_{GS}} \qquad (11)$$

Putting $kT/C_{GS}=59×10^{-12}$ V$^2$ in Eq. (11) the mean square resistance noise of our 3,5 kΩ resistor is: $<(\Delta R_{ch})^2>=4.3×10^{-6}$ Ω$^2$. This is a noise of 2.1 mΩ$_{rms}$, thus a relative noise $\Delta R_{ch}/R_{ch}=5.9×10^{-7}$ (≈0.6 ppm). Although not far from the $\Delta R/R≈10^{-7}$ value we gave in the Introduction, this *Lorentzian FIRN in TE* is higher because our technology to give rigidity to the n-GaAs layer of our resistor is a "very noisy" one. Replacing its p$^+$-GaAs substrate by a lightly doped p-GaAs substrate (see below) or by a SI-GaAs one, the backgating effect on the n-GaAs layer would be lower, thus decreasing this Lorentzian FIRN whose existence in the channel seems undeniable by considering that thermal equipartition sets the $kT/C_{GS}$ noise of its $C_{GS}$ in TE.

Eq. (11) that links mean square values coming from Lorentzian spectra with equal cut-off frequency $f_c$ also should link at each frequency the spectral density $S_R(f)$ in Ω$^2$/Hz with its cause: the spectral density $S_V(f)$ in V$^2$/Hz given by Eq. (5). From the flat spectral density $S_{V0}=4kTR_{GS}$ V$^2$/Hz at low frequencies, the low frequency density $S_{0R}$ of resistance noise in the channel will be:

$$S_{OR} = \frac{R_{ch}^2}{4\Phi^2}\left(\frac{0.15}{0.75}\right)^2 \times 4kTR_{GS} = 2.7 \times 10^{-5} \ \Omega^2/Hz \qquad (12)$$

Thus, our resistor of Fig. 5 in TE will offer a mean resistance $R_{ch}$=3.5 kΩ with a small resistance noise of null mean value (like its source: the $kT/C_{GS}$ noise) and Lorentzian spectrum given by:

$$S_R(f) = \frac{S_{0R}=2{,}7\times 10^{-5}}{1+\left(f/f_c\right)^2} \ \Omega^2/Hz \qquad (13)$$

Dividing Eq. (13) by $(R_{ch})^2$ to have the normalized fluctuations of Eq. (2), the flat region of this spectrum gives: 2.2×10$^{-12}$ Hz$^{-1}$ (-117dB). From this value and its $f_c$=0.1Hz we have drawn its asymptotic Bode plot labelled (B) in Fig. 2. This is the normalized Lorentzian of FIRN in TE of the resistor of Fig. 5 that would contain N=10$^{10}$ free carriers provided that its 10$^{10}$ donors in its n-GaAs layer were fully ionized. This Lorentzian $S_R(f)$ in TE will be a key reference to compare other resistance noise spectra that we will find in this resistor out of TE. It is worth mentioning the proximity of this graph to point M of Fig. 2, which was drawn for the normalized fluctuations of resistance with 1/$f$ spectrum of a resistor with equal number of free carriers in its channel (N=10$^{10}$) made from a "noisy technology" giving rise to $\alpha_H$=10$^{-3}$. At the end of this work we will compare this graph (B) of the resistor in TE (Lorentzian) with its normalized FIRN out of TE (1/$f$ noise) when a current $I_{conv}$=150μA is injected to observe its $S_R(f)$ by its excess noise.

To see an image of the Lorentzian FIRN of Eq. (13) by a spectrum analyzer we decide to convert this $S_R(f)$ into noise voltage by injecting in our 3.5kΩ resistor enough $I_{conv}$ so as to surpass its Johnson floor of $4kTR_{ch}=5.8\times10^{-17}$ V²/Hz. From Eq. (1) the minimum $I_{conv}$ we would need to observe a Lorentzian spectrum of noise voltage $\Delta V$ over such a floor (i. e. to see *an image of $S_R(f)$* as an excess noise *emerging over $4kTR_{ch}$*) should be:

$$(I_{conv})^2 \geq \frac{4kTR_{ch}}{S_{0R}} \Rightarrow I_{conv} \geq 1.5\mu A \qquad (14)$$

Apparently, a 100 times higher $I_{conv}=150\mu A$ should give the excess noise density of Fig. 6 that floating at 40dB over the Johnson floor, would give a "clean image" of the Lorentzian spectrum of Eq. (13). From its flat region lying 40dB over the floor, this spectrum and its floor will meet at $f=100f_c$ (see Fig. 6). Since the heating power $P_Q=(I_{conv})^2\times R_{ch}=79\mu W$ seems low for our 1mm long device onto a thick GaAs substrate sinking the generated heat, its Johnson floor roughly should be that of $T=300K$, thus: $4kTR_{ch}\approx 5.8\times10^{-17}$ V²/Hz.

**Figure 6.** Lorentzian spectrum of excess noise that $I_{conv}=150\mu A$ would reveal from the resistance noise of the resistor of Fig. 4 if its presence in the channel would not disturb the TE of this device (*theoretical wish*).

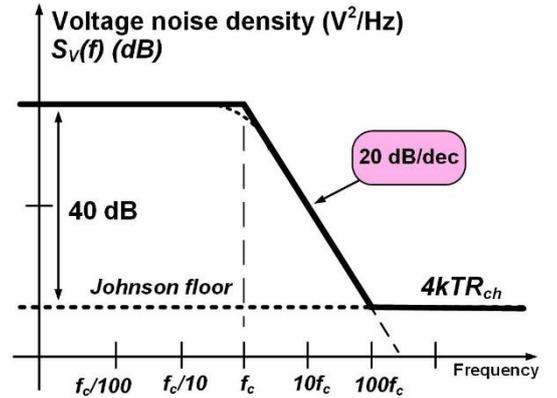

Setting $I_{conv}=150\mu A$ in our resistor of Fig. 5 and looking for its excess noise in the spectrum analyzer we would see its Johnson floor, but NOT the Lorentzian excess noise of Fig. 6. The excess noise we would observe would look like the 1/*f* excess noise of Fig. 1 described by Eq. (2). The reason for this disappointing result is that the Lorentzian excess noise drawn in Fig. 6 only is this *theoretical wish*: $(I_{conv})^2$ times the resistance noise of the resistor of Fig. 5 *in TE*, whereas the observed spectrum is $(I_{conv})^2$ times the resistance noise of the resistor *out of TE* due to $I_{conv}$ itself and this new FIRN does not follow Eq. (13) as we comment in the next paragraph from our Fluctuation-based noise model [4, 5].

As soon as $I_{conv}\neq 0$ is set in its channel, the resistor no longer is in TE. The way fluctuations of energy in $C_{GS}$ relaxed in TE with a single time constant $\tau_{gs}$, no longer holds and they pass to relax with a wide set of time constants along the channel that covers many decades of frequency [6]. To see a Lorentzian excess noise like that of Fig. 6 in a resistor biased by $I_{conv}\neq 0$ we have to resort to a striking property of the FIRN: its trend to increase as the device causing it is shrunk. For this purpose, Fig. 7 shows the same channel of the resistor of Fig. 5 but with two strip-like gates of $p^+$-GaAs replacing its 1mm long and continuous gate due to its p$^+$-GaAs substrate. In this way our resistor that was a normally-ON JFET with $R_{ch}$=3.5kΩ, becomes a double gate JFET with the same conductive channel coming from the series connection of two JFETs of length L/2. Equally valid is to consider our new resistor as the series connection of 20 short JFETs, each L/20=50µm long, where two of them are noisy due to their floating gates and eighteen of them are noiseless (the gateless ones). This picture suggesting that our resistor could be less noisy because only one tenth of its channel has FIRN, is very educational to learn about resistance noise. The reason is that the new device of Fig. 7 in TE is a resistor of mean resistance $R_{ch}$=3.5kΩ like that of Fig. 5, but *two times noisier*. And moreover: under $I_{conv}$=150µA, its excess noise between terminals will be two times the Lorentzian one of Fig. 6 that the resistor of Fig. 5 is unable to show while its $I_{conv}$=150µA is activated. Due to this ability that we will justify soon let us consider the resistor of Fig. 7 as a 3.5kΩ *calibrator* for resistance noise measuring systems.

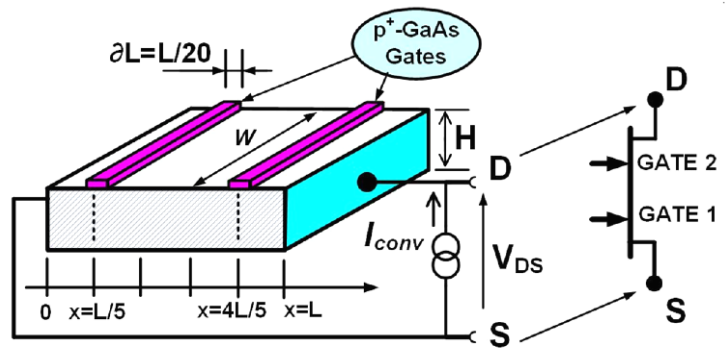

**Figure 7.** Planar resistor with two strip-like gates on top designed to produce Lorentzian excess noise spectra in its channel from the *kT/C* noise of its stripped BSCR under such gates (see the text).

The double resistance noise of this calibrator is easy to explain from the twenty times higher thermal noise of the input capacity $C_{GS}$/20 of each of its short JFETs at x=L/5 and x=4L/5. This capacity of each strip diode shunted by a twenty times higher resistance 20$R_{GS}$ will give the same cut-off frequency $f_c$ of Eq. (7) for their Lorentzian thermal noises in TE. Thus there is 20 times higher thermal noise modulating the 20 times lower resistance ($R_{ch}$/20=175 Ω) of the channel of each short JFET. This means that each strip gate of 20 times lower area than L×W=0,1 mm$^2$ (the gate area of Fig. 5) is modulating 20 times more deeply its portion of the channel of 175Ω. In this way, each short JFET of Fig. 7 in TE sets an amount of resistance noise *in its* 50µm *channel* that equates the resistance noise that the 1mm long gate of Fig. 5 uniformly creates in its 1mm long channel.

This shows why FIRN is "the noise of the small layouts" as we wrote in the next-to-last paragraph of the Introduction with regard Hooge's formula predicting this feature. Given that these two short FETs generate resistance noise from the thermal noise of their independent diodes far apart in the channel (at $x_1$=200µm and at $x_2$=800µm) we will take them as *uncorrelated* noises that must be added *in power* to give the whole resistance noise of the channel. This means that the spectrum of FIRN in the channel of the calibrator of Fig. 7 in TE will be twice the spectrum of FIRN in the resistor of Fig. 5 in TE. Doubling Eq. (13) the resistance noise in the channel of the resistor of Fig. 7 becomes this one:

$$S_R(f) = \frac{2S_{0R}=5{,}4\times 10^{-5}}{1+\left(f/f_c\right)^2} \quad \Omega^2/Hz \qquad (15)$$

Besides its 3dB higher resistance noise in TE, the calibrator of Fig. 7 differs from the resistor of Fig. 5 concerning the excess noise it will show for $I_{conv}\neq 0$. We refer to the fact that the FIRN in the channel of Fig. 7 is not uniformly generated as it is in the channel of Fig. 5, but *localized* in two small regions around x=L/5 and x=4L/5. Due to this feature, the presence of $I_{conv}$ in the channel of Fig. 7 (that only gives a voltage drop $V_{DS\partial}\approx 25$mV in each FET) is a *low-level disturbance* that barely changes the "environment" of the channel region where resistance noise was generated in TE under $V_{DS\partial}=0$. To show why this feature avoids the 1/*f* excess noise that will appear in the resistor of Fig. 5 under $I_{conv}=150\mu A$, let us study next the logarithmic voltage divider that appears in the resistor of Fig. 5 when it is biased by enough current $I_{conv}$.

**IV- A logarithmic converter from a long channel FET**

Recalling that the Lorentzian spectrum of Fig. 6 corresponds to the excess noise that we should find in the resistor of Fig. 5 if it remained in TE while $I_{conv}$ is set, a likely reason why we do not find this spectrum is that using $I_{conv}$ to reveal as excess noise a small resistance noise $\Delta R_{ch}/R_{ch}\approx 10^{-7}$ is a *thermally aggressive* method. By "aggressive" we do not mean a high heating of this 1mm long device. What we mean is that $V_{DS}=R_{ch}\times I_{conv}=525$mV, the voltage drop existing between terminals of this device biased by $I_{conv}$, means that *it is far from TE* despite its negligible temperature change. Considering that the thermal voltage unit at room *T* is $V_T\approx 25{,}9$ mV, voltages between terminals close to $V_T$ would allow thinking of this device as *"not far from TE"*, but voltages surpassing 20 thermal units should be taken with care. Note that $V_{DS}=525mV\approx 20{,}3V_T$.

To face this situation let us go back to Fig. 5 presenting the resistor as a voltage divider driven by the voltage $V_{DS}$ between its D and S terminals and whose output is the voltage $V_{FL}$ the Gate acquires respect to its S terminal due to $V_{DS}\approx 20.3V_T$. From the coupling between Gate and channel provided by the $p^+$-n diode let us obtain $V_{FL}$ by taking the Gate as *an equipotential region* not only due to its high conductivity, but also to the tiny currents associated with the gate of a GaAs FET like this one. If Si-based JFETs using $p^+$-n junctions with barriers close to $q\Phi\approx 1{,}1$eV show gate currents under the nA at room *T*, the gate of our resistor using a $p^+$-n GaAs junction with $q\Phi\approx 1{,}3$eV should show currents under the pA. After this brief comparison of reverse saturation currents of GaAs and Si junction diodes, let us show what happens when $I_{conv}=150\mu A$ is set in this resistor.

Since $V_{DS}=525$mV is well below the pinch-off voltage of the channel of this resistor ($V_p\approx 46$V), the cross section of its channel along x in Fig. 5 will be quite the same, thus giving rise to a *linear drop of* $V_{DS}$ along it. Thus, the voltage within the channel at x=L/2 would be: $V_{DS}/2$=262.5 mV that is a first value we will use for $V_{FL}$ to launch a reasoning seeking its actual value. If $V_{FL}$ was $V_{DS}/2$ mV, the junction voltage in the p$^+$-n diode would go from $V_J(x)$=+262.5 mV at x=0 (next to the S terminal) to $V_J(x)$=-262.5 mV at x=L (next to the D terminal). A line along W at x=L/2 where $V_J(x)$=0 would separate the lower half area of the diode *under forward bias* from the upper one *under reverse bias*. From the exponential *i-v* characteristic of the p$^+$-n diode, the current going from the floating gate towards the channel through its lower half area would be higher than the current going from the channel to this gate through its upper half area. This would reduce the gate voltage down to a voltage $V_{FL}$ equating these two currents in order to make null the *net current* leaving or entering the floating Gate.

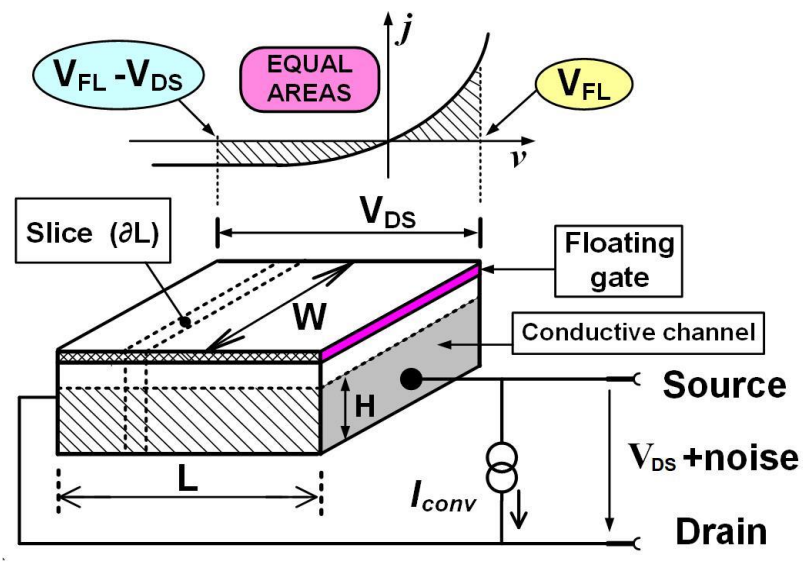

**Figure 8.** Geometrical way to obtain the voltage $V_{FL}$ the floating Gate has to acquire in Fig. 5 in order to balance the current it extracts from the channel and the current it injects to the channel.

Fig. 8 shows the way $V_{FL}$ is found by means of the i-v curve of the p$^+$-n diode. Since all the slices of Fig. 8 have the same width W, the $V_{FL}$ voltage we are looking for appears by placing on the voltage axis of the i-v curve a segment of $V_{DS}$ volts, in such a way that the two dashed areas, which are proportional to the forward and reverse current entering and leaving the floating Gate, become equal. The *i-v* and *j-v* characteristics we mean are these:

$$i = I_{sat} \times \left[exp\left(\frac{v}{mV_T}\right) - 1\right] \; or \; j = J_{sat} \times \left[exp\left(\frac{v}{mV_T}\right) - 1\right] \quad (16)$$

where *m* is the ideality factor and $V_T=kT/q$ is the thermal voltage unit. Nulling the integral of this i-v or j-v from $V_{FL}$ (the highest forward bias of the p$^+$-n diode next to the S terminal in Fig. 5) to ($V_{FL}$-$V_{DS}$) (the highest reverse bias of the diode next to the D terminal) we obtain this logarithmic transfer function:

$$V_{FL} = mV_T Ln\left[\frac{\frac{V_{DS}}{mV_T}}{\left(1-exp\left(\frac{-V_{DS}}{mV_T}\right)\right)}\right] \approx mV_T Ln\left[\frac{V_{DS}}{mV_T}\right] \; for \; \frac{V_{DS}}{mV_T} > 4 \quad (17)$$

Using m≈1 to simplify and high $V_{DS}$ values ($V_{DS}>>V_T$) the number of thermal units $V_T$ that the floating Gate acquires is the natural logarithm of the number of $V_T$ units along the channel. For $V_{DS}>4V_T$ (100 mV at room $T$) this means that as $V_{DS}$ rises linearly, $V_{FL}$ rises logarithmically. For $V_{DS}$=525mV we have: $V_{FL}≈3V_T$=78 mV and for $V_{DS}$=1V it would be: $V_{FL}≈$95mV. When $V_{DS}$ is small ($V_{DS}<<V_T$) the voltage of the floating Gate tends to be $V_{FL}≈V_{DS}/2$. Therefore, setting $I_{conv}$=150µA along the channel of Fig. 5 produces a *non-uniform biasing* (NUB) of its $p^+$-n diode that goes from $v=V_{FL}$=+78mV (+3$V_T$ forward bias) next to S terminal, to $v=V_{FL}-V_{DS}$ ($v$=-447mV=-17,3$V_T$) next to D terminal. This NUB along x suggests the set of slice-like FETs of Fig. 8, each creating FIRN in the channel from a thermal noise with its own relaxation dynamics or time constant $\tau_{gs}(v)$ [4, 5].

This NUB decomposes the $p^+$-n diode between channel and substrate into a continuous set of narrow strip-diodes, each biased with ($V_{FL}$-$v$) volts, thus giving rise to a continuous set of Lorentzian spectra of electrical noise along x, which in turn gives rise to the generation of a continuous set of Lorentzian spectra of FIRN in the slices that form the channel following Fig. 8. From the exponential change of the $R_{GS}$ of a $p^+$-n diode with its bias voltage $v(x)$, the cut-off $f_c$ of these spectra of FIRN spectra will cover *nine decades of frequency*. From previous words we could repeat the continuous treatment done in [6] to obtain the 1/$f$ FIRN created by the current $I_{conv}$ but this could hide relevant notions. This is why we propose to use the calibrator of Fig. 7 to learn about this resistance noise that has puzzled scientists, not only because the 1/$f$ spectrum of excess noise it gives but mainly concerning its genesis. For this purpose, next Section shows a few notions on the way the Lorentzian noise voltage of a diode is shaped by the dc voltage that a small dc current sets between its terminals [12, 13].

**V- Noise tunability in junction diodes and FIRN tunability**

From the noise viewpoint, the NUB of the $p^+$-n diode under $I_{conv}$≠0 has two effects on the input admittance of the long FET of Fig. 5 whose value in TE was $C_{GS}$ shunted by $R_{GS}$. The first effect concerns the set of capacities that should replace $C_{GS}$ obtained when the thickness $t_{back}$ of its BSCR was the same along the *equipotential* channel. It is worth noting that a single capacity like $C_{GS}$ can be found between *two equipotential conductors* like the $p^+$-substrate and the channel in TE, but when the later loses this property due to $I_{conv}$≠0, we have to take with care the *set of capacities* to be considered. Because for $I_{conv}$≠0 there is a linear voltage drop along x, the BSCR thickness $t_{back}(v)$ becomes position-dependent, see Eq. (10). In this case the "capacity between Gate and channel" is not defined because the energy an electron needs to pass between the non-equipotential channel and the $p^+$-gate to create a fluctuation of energy accordingly to [4, 5] varies with the position x along the channel. Properly speaking, there will be a set of *differential capacities at each position x*. However, the changes of voltage in the junction along the channel are low enough so as to consider that the capacity per unit area of the diode along x roughly is equal to $C_{GS}/(L×W)$ where $C_{GS}$=70pF is the capacity of the diode in TE. From Φ=1,3V, Eq. (9) and Eq. (10) we can see that the change of the junction capacity going from (Φ-78mV) to (Φ+447mV) is small as compared with the second effect we are to consider in next paragraph. Thus, the capacity of each slice-like junction diode drawn by dotted lines in Fig. 8 roughly will be: $\partial C_{GS} \approx C_{GS} \times (\partial L/L)$.

The second effect to consider is the *exponential* way this NUB of the $p^+$-n diode affects the resistance $\partial R_{GS}(v)$ that shunts its local capacity $\partial C_{GS}$ at each x, thus setting the cut-off frequency $f_c(v)$ of the thermal noise producing FIRN at position x in the channel. This $\partial R_{GS}(v)$ together with $\partial C_{GS}$ set the *relaxation dynamics* for the *dissipation* of energy fluctuations appearing in the capacity $\partial C_{GS}$ of each local diode [4, 5]. Differentiating Eq. (10) the dynamical resistance of the channel-substrate diode as a function of its bias voltage *v* at each point is this function:

$$R_{GS}(v) = \left(\frac{di}{dv}\right)^{-1} = \frac{mV_T}{I_{sat}} \times exp\left(\frac{-v}{mV_T}\right) = R_{GS} \times exp\left(\frac{-v}{mV_T}\right) \quad (18)$$

In contrast with the weak dependence of $C_{GS}(v)$ with v, Eq. (18) shows an *exponential change for the resistance setting the cut-off frequency $f_c$ of the FIRN* at each position *x* in the channel. This leads to see our resistor as the connection in series of those short FETs of length $\partial L$ and width W (slices) shown in Fig. 8. For m=1 and because the junction area of each slice is $\partial L/L$ times the area W×L, the set of resistances shunting the input capacities of these slices of the channel of our FET of Fig. 5 with $V_{DS}$=525mV will go from $\partial R_{GS}$=(L/$\partial L$)×exp(-3)×22.7GΩ next to the S terminal where v=+3$V_T$ to $\partial R_{GS}$=(L/$\partial L$)×exp(17.3)×22.7GΩ next to the D terminal where v=-447mV=-17.3$V_T$. Multiplying each $\partial R_{GS}$ by its shunting $\partial C_{GS}$, their inverse "area factors" $\partial L/L$ and L/$\partial L$ mutually cancel and the set of time constants thus obtained goes from 1.59×exp(-3)=0.08s ($f_c$=2Hz at terminal S) to exp(17.3)×1.59=5.2×10$^7$s ($f_c$=3×10$^{-9}$ Hz at terminal D). This means more than *nine decades* of $f_c$ values for the Lorentzian noises of this set of "slice-like" FETs that form the whole channel of the resistor.

Thus, a "moderate" $V_{DS}$=525mV between terminals of our resistor gives a set of Lorentzian spectra of FIRN in its channel with $f_c$ sweeping more than eight decades of frequency under the Hz. In contrast with well-known works on 1/*f* noise like [14] our *non-rigid channel* approach is a radical proposal that does not need fluctuations of conductivity to explain resistance noise in resistors. Added to this, our model predicts a broad, continuous set of Lorentzian resistance noises in the channel of a resistor due to its own biasing and the effect that creates this continuous set of Lorentzian terms whose $f_c$ cover many decades of frequency is the familiar field effect of devices like JFETs, MOST and vacuum tubes. Once the presence of this broad set of FIRN terms in the channel of our resistor has been shown, this question arises: Are they *properly weighted* so as to synthesize a 1/*f* excess noise spectrum? This "proper weight" means that the amplitude of each Lorentzian term is inversely proportional to its $f_c$ [10, 14], as Eq. (5) predicts when $\partial C_{GS}$ roughly is constant while $\partial R_{GS}$ varies following Eq. (18) [12, 13].

Thus, the answer to previous question is "yes" as shown in [6] by a rather numerical reasoning we are replacing by a more enlightening way to understand FIRN with the help of Fig. 7. We are doing this because the unawareness of FIRN that experts in the field seem to show, perhaps guided by the notion of this noise coming from the "unlimited bulk" of a "sample". After reading [15] published some months after [6] we noticed its authors on our paper showing that the 1/$f$ excess noise was a combined effect coming from the familiar field effect and the disturbance caused by the method used to reveal resistance noise as excess noise. Since [6] solved the endless conflict between a school of thought assigning 1/$f$ noise to fluctuations in mobility ($\Delta\mu$ noise) and the other school assigning it to carrier number fluctuations ($\Delta N$ noise) [14] we hoped that they would understand our proposal. However, the scarce interest they showed on the subject led us to realize that [6] was hard to understand, thus requiring the set of graphical and intuitive notions we are giving here.

**VI- Lorentzian FIRN from small devices**

Going back to the calibrator of Fig. 7 let us activate its $I_{conv}$=150µA to obtain its excess noise between terminals. Taking one of its two noisy, short JFETs of 50µm channel, we can observe that its drain-source voltage drop due to $I_{conv}$ only is ≈1$V_T$ because $V_{DS\partial}$=525mV/20≈26mV. Thus, each JFET is a small device *not far from TE* despite its $I_{conv}$. From the linear drop of $V_{DS}$ along x the voltages of the channel at x=L/5 and x=4L/5 would be: $v_1$=105mV and $v_2$=420mV. From Eq. (17) with the small voltages $V_{DS\partial}$≈26mV of these short FETs, the voltages of the floating Gate1 and Gate2 will track them: $V_{FL1}$≈$v_1$ and $V_{FL2}$≈$v_2$. Thus, each short FET of Fig. 7 remains "near TE" despite the disturbance of $I_{conv}$=150µA along its short channel and its spectrum of FIRN will be close to its Lorentzian one in TE. The name calibrator given to this device is due to this ability to give a Lorentzian spectrum of excess noise for $I_{conv}$≠0 in its channel. Let us show why this is so by considering that this excess noise comes from FIRN created in its channel by its two gates left floating, thus from resistance fluctuations *localized* around x=L/5 and x=4L/5 that a spectrum analyzer cannot distinguish from resistance noise uniformly generated along the channel (see the end of Section II).

Since each short FET of Fig. 7 is in "quasi-TE" despite $I_{conv}$=150µA, their input admittance will be $C_{GS}$/20 shunted by 20$R_{GS}$. Thus, the cut-off frequency $f_c$ of its Lorentzian resistance noise will be: $f_c$=0.1 Hz as it was in TE (and as it is in the resistor of Fig. 5 in TE). As we have shown in Section III, the FIRN of this calibrator of Fig. 7 in TE is twice the resistance noise of the resistor of Fig. 5 in TE and it is given by Eq. (15) coming from the sum in power of the FIRN in each channel of its two short FETs. This duplication shows the striking increase of FIRN as the device shrinks because from two small gates whose areas only sum one tenth of the gate area of Fig. 5, the channel of Fig. 7 becomes two times noisier with the same geometry and resistance $R_{ch}$=3.5kΩ than that of Fig. 5.

With its two gates left floating, the excess noise of the calibrator of Fig. 7 biased by $I_{conv}$=150µA would be spectrum A of Fig. 9. Shorting its Gate1 to its S terminal while its Gate2 is left floating, or shorting its Gate2 to its S terminal while its Gate1 is left floating, it would show the spectrum B of Fig. 9 because one of its two sources of FIRN (Gates) has been "silenced". Shorting both gates to its S terminal it would show its floor of Johnson noise for $I_{conv}$=150µA that could be compared with its floor for $I_{conv}$=0 to track a possible heating. These results that could be useful to build Lorentzian calibrators of resistance noise also mean that Lorentzian spectra of excess noise usually assigned to fluctuations in the number of carriers (ΔN noise [15]) *could equally come from dielectric relaxations* having nothing to do with carrier number fluctuations.

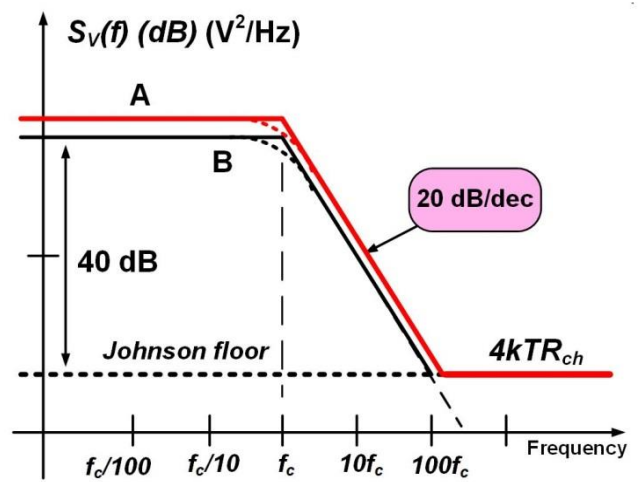

**Figure 9.** Lorentzian spectra of excess noise that $I_{conv}$=150µA would produce from the FIRN that exists in the channel of the resistor of Fig. 7 while $I_{conv}$ is set in its channel (real spectrum, graph A). Graph B representing any of the two components of graph A, is equal to the spectrum of Fig. 6 (see the text).

Therefore, Lorentzian excess noise can come from fluctuations in the cross section of channels caused by fluctuations of electric field in BSCRs linked with interfaces and surfaces in their vicinity or linked with dislocations and defects embedded in the channel that use to be surrounded by SCRs. Using this feature, the calibrator of Fig. 7 is emulating Lorentzian excess noise that could be equally assigned to an *unreal carrier trap in its bulk* of material. And moreover: the "existence" of this inexistent trap would seem backed by the activation energy ΔE≈qΦ eV we would obtain from the thermal evolution of its $f_c$ [16] (e. g. by an Arrhenius plot). From the GaAs bandgap ($E_g$=1.42 eV) people unaware of the role of the *device* in these measurements could assign this FIRN with ΔE≈1,3 eV to deep traps "in the bulk" of this device. Although these dielectric relaxations that mimic carrier traps could have to do with "deep levels near $E_g$/2" found in GaAs devices onto SI-GaAs substrates, this topic falls out of the scope of this paper and we will leave it just after showing the main lesson that the calibrator of Fig. 7 offers when its two floating gates are shorted as shown in Fig. 10.

To deal with diodes far away from TE like those we will meet in the device of Fig. 10, let us consider the origin of the *kT/C* noise of a two-terminal device like a junction diode. Accordingly to [4, 5], this noise voltage comes from a huge rate of tiny fluctuations of electrical energy occurring in its capacity *C*. The mean energy of these fluctuations is $U=q^2/(2C)$ J, the energy that the electronic charge *q* displaced between terminals sets in *C*. For C=3,5pF (the capacity $C_{GS}/20$ of the input admittance of each short FET of Fig. 7) this tiny fluctuation is: $U \approx 2.3 \times 10^{-8}$ eV, thus *less than 1 ppm* of the thermal unit of energy *kT*=25,9 meV at room *T*. This explains why the rate of such passages is so high that noise voltage looks like continuous. Since these fluctuations form an impulsive charge noise in *C* due to *displacement currents* producing *shot noise*, the noise voltage of a junction diode in TE should come from its two equal, but opposed, saturation currents $I_{sat}$ crossing its SCR for *v*=0.

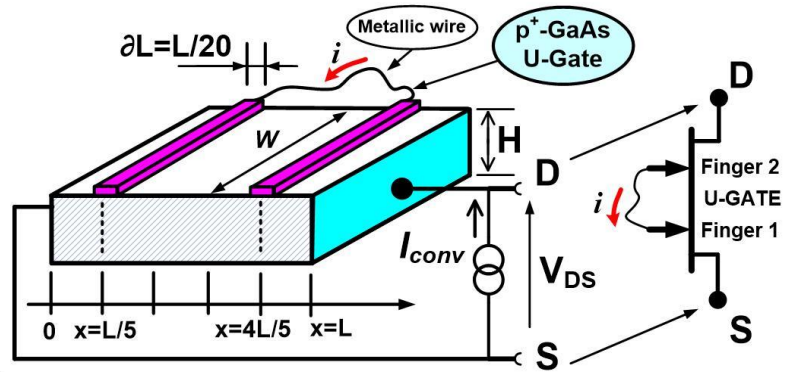

**Figure 10.** Planar resistor with two strip-like gates on top wired as a U-Gate that generates resistance noise in the channel under its fingers (see the text).

These $I_{sat}$ across a SCR empty of carriers must be *displacement currents* (not conduction ones) producing electrical noise [4], where we also wrote that the circuit simulator PSPICE does not consider this fact because it only assigns shot noise to the *net current* of the diode that is null in TE. Thus, junction diodes in TE are noiseless for PSPICE but not for our model where their two $I_{sat}$ that mutually cancel *on average* but not at each instant of time are taken as uncorrelated displacement currents. This means that their shot noise must be added in power to obtain the density $S_I(f)$ of shot noise of the diode in TE that becomes twice the density $S_{Ishot}(f)=2qI_{sat}$ A²/Hz of each $I_{sat}$. Thus, $S_I(f)=4qI_{sat}$ A²/Hz driving $Y(j\omega)$ (the admittance of the diode) produces its *noise voltage* that can be taken as Johnson noise associated to its dynamical $R_{GS}$ or as the $kT/C_{GS}$ noise of its capacity $C_{GS}$ due to thermal equipartition, recall Eqs. (5), (6). From $R_{GS}$ and $C_{GS}$ in parallel we obtain the admittance $Y(j\omega)$ converting the noise current density $S_I(f)=4qI_{sat}$ A²/Hz into the noise voltage density between terminals we can measure. This gives:

$$S_V(f) = 4qI_{sat}\left|\frac{1}{Y(j\omega)}\right|^2 = \frac{4kTR_{GS}}{1+\left(f/f_c\right)^2} \quad V^2/Hz \qquad (19)$$

Eq. (19) that is previous Eq. (5) shows that the spectrum of noise voltage of the diode in TE is the Johnson noise ($4kTR_{GS}$ V²/Hz) of a resistor of resistance $R_{GS}$ filtered by $C_{GS}$ (noise picture where resistors generate noise and capacitors do not generate it). Since the integration of Eq. (19) gives the $kT/C_{GS}$ noise of $C_{GS}$ obeying thermal equipartition in TE, recall Eq. (6), Eq. (19) also shows the $kT/C_{GS}$ noise of $C_{GS}$ (equipartition) that is shaped by the relaxation dynamics set by $R_{GS}$ and $C_{GS}$ working together [4, 5]. Replacing $R_{GS}$ and $C_{GS}$ by $20R_{GS}$ and $C_{GS}/20$ in Eq. (19) the noise voltage in the capacity of each short FET of Fig. 7 is:

$$S_V(f) = \frac{4kT(20R_{GS})}{1+\left(f/f_c\right)^2} \quad V^2/Hz \qquad (20)$$

Eq. (20) that also would appear from the shot noise $S_I(f)=4q(I_{sat}/20)$ A²/Hz of the two opposed $I_{sat}/20$ of the strip diodes below the fingers of Fig. 7 and their 20 times lower admittance means that these diodes are noisy in TE despite the model of PSPICE violating equipartition [4]. Thus, let us use our fluctuation-based model to answer this question: What will be the spectrum of noise voltage in one of these finger diodes when it becomes reverse biased by a current setting its reverse voltage to $v=-4V_T$? The answer to this question from the dissipation-based model for electrical noise in use today is not easy, but following [4, 5] the answer appears from these two actions modifying Eq. (20):

**a)** *Shot noise evaluation*: divide by two Eq. (20) because the shot noise density of this diode under $4V_T$ reverse voltage must be: $S_{Irev}(f)≈2qI_{sat}$. This is so because one of its $I_{sat}$ vanishes under this reverse voltage while the other remains.

**b)** *Spectrum tuning*: multiply the numerator of Eq. (20) by $exp(4)≈50$ to account for the 50 times higher $R_{GS}$ of the diode under $v=-4V_T$, see Eq. (18) and divide its cut-off frequency $f_c$ by $exp(4)≈50$ because its capacity $C_{GS}/20$ for $v=0$ roughly is equal to its new capacity under $v=-105mV$.

These two actions summarize the tuning of the noise voltage confirmed in broad area, silicon Schottky diodes [12, 13]. These devices were measured due to their low barrier Φ and broad area helping to obtain low enough dynamical resistances $R_{GS}$ allowing to keep loading effects of the surrounding electronics at an acceptable level. Otherwise, the noise of the diode is masked by such loading. Thus action **a)** takes into account the amount of shot noise in the diode whereas action **b)** sets its relaxation dynamics for the conversion of its noise current into noise voltage between its terminals [4, 5]. By the way: the noise voltage PSPICE gives for a diode under this bias "converges" towards the noise we obtain from actions **a)** and **b)** because $I_{net}$ tends to be: $-I_{sat}$ and this program computes very well both the capacity and the dynamic resistance of the diode.

Since Eq. (20) for the noise voltage in TE of each junction diode under the strip gates of Fig. 7 is 20 times Eq. (5) and it only produces FIRN in its channel of length L/20, the FIRN spectrum it adds to the 1 mm long channel of the calibrator will be graph B of Fig. 9. Thus, the evolution of Eq. (20) following actions **a)** and **b)** dictated by the bias conditions of these diodes in Fig. 10 will allow to track the evolution of their excess noise. In this way, previous action **a)** that was: "divide by two Eq. (20)" would translate into a vertical drop of 3dB in graph B of Fig. 9 and previous action **b)** means that the graph B just obtained from action **a)** must be displaced towards low frequencies *along a line with 1/f slope* (10dB/decade) to reach its new $f_c$. Once this has been realized we have the tools to follow next Section showing how FIRN that obeys Eq. (2) is born in the channel of a resistor from the apparently "*linear noise converter*" that $I_{conv}$ looks like in Eq. (1).

## VII- Synthesis of 1/f excess noise in resistors from their FIRN

To gain some insight into the new device of Fig. 10 whose fingers form an *equipotential U-gate*, let us obtain its FIRN in TE to be compared with that of the calibrator of Fig. 7. Due to the metallic wire the capacity of the $p^+$-n junction under the U-Gate thus formed is $C_{GS}/10$. Thus, the $kT/C_{GS}$ $V^2$ producing FIRN in the whole channel of Fig. 5 has to be replaced by $kT/(C_{GS}/10)$ $V^2$ producing FIRN under each of the fingers of this U-Gated resistor. In this way the mean square noise voltage creating FIRN under each finger of Fig. 10 is *half the value* found for each finger of Fig. 7, whose fingers worked *independently* as uncorrelated sources of FIRN. The result is that the $kT/(C_{GS}/10)$ $V^2$ producing FIRN under the fingers of this U-Gated resistor is half the $kT/(C_{GS}/20)$ $V^2$ under the fingers of Fig. 7. Apparently, the FIRN in the channel of the calibrator of Fig. 10 in TE would be half the FIRN in the channel of the calibrator of Fig. 7 in TE.

However, this is not so because the two fingers of the calibrator of Fig. 10 shorted by the metallic wire will work *synchronously* in TE, thus creating *equal modulations at each instant of time* in the 50µm channel of each short FET. This means that the Lorentzian spectra of FIRN in the channels of these short FETs must not be added in power like the uncorrelated FIRN of the fingers of Fig. 7. They must be added "*in value*" and because the sum (a+a=2a) of two equal values gives 2 times higher power ($4a^2$) than its sum in power ($a^2+a^2$)=$2a^2$, the 2 times more power of this sum in value compensates for the 2 times lower mean square voltage generating FIRN under the fingers of Fig. 10. Thus, the spectrum of FIRN in TE of the calibrator of Fig. 7, that multiplied by $(I_{conv})^2$ gave graph A of Fig. 9, also is the spectrum of FIRN in TE of the calibrator of Fig. 10.

In summary: with regard FIRN in TE between their S and D terminals, the two times higher modulation (mean square) of the channels of the FETs of Fig. 7 is counterbalanced by the fact that such modulations are uncorrelated and must be added in power to obtain the total FIRN in the channel of this calibrator. By contrast, the two times lower modulation (mean square) of the channels of the FETs of Fig. 10 is counterbalanced by their sum in value due to their coherent nature. All in all, the FIRN of the calibrators of Figs. 7 and 10 in TE would have the same spectrum that multiplied by $(I_{conv})^2$ would give graph A of Fig. 9. Thus, an external observer able to measure the FIRN between the S and D terminals of these calibrators *while keeping them in TE*, "would not know" if there is or there is not, a metallic wire connecting their two finger-like gates. This is a cogent result because the net current of this wire is null in TE and cutting this wire in the device of Fig. 10 should not affect the spectrum of FIRN measured in its channel.

Thus, a non-disturbing measurement of their FIRN in TE would not tell us which calibrator is being used. Only from those internal details we know, we can speak about FIRN generated at specific regions of the 1 mm long channel of the calibrator or about the coherent or incoherent sum of resistance noises generated at different regions. Measuring the calibrators of Figs. 7 and 10 by an "ohmmeter" we would find them as two similar resistors of $R_{ch}$=3.5 kΩ between its terminals (mean value). Measuring their Johnson noise between terminals (thus in TE) we would find them as two undistinguishable resistors of $R_{ch}$=3.5 kΩ generating similar Johnson noise. But putting them out of TE by injecting similar currents $I_{conv}$=150µA in their channels to measure their excess noise, we would find a striking difference. Whereas the resistor of Fig 7 would show the Lorentzian excess noise with $f_c$=0.1 Hz shown by graph A of Fig. 9 as explained, the resistor of Fig. 10 would show *two Lorentzian spectra* of excess noise. One of these spectra would be a "hot" spectrum with higher cut-off frequency ($f_{c1}$=2$f_c$) and lower amplitude than graph A of Fig. 9 and the other would be a "cold" spectrum with much lower cut-off frequency ($f_{c2}$=2,6×10$^{-7}$ Hz) and much higher amplitude.

To explain the origin of this amazing splitting of FIRN let us advance that when $I_{conv}$≠0 is set in the calibrator of Fig. 10, the coherent sum of FIRN in its channel no longer occurs because *the fluctuations* causing the *kT/C* noises of the diodes under their fingers *no longer are synchronized* accordingly to [4, 5]. With regard the input admittance of the U-gated JFET of Fig. 10 in TE it would be its $C_{GS}$/10 shunted by its 10$R_{GS}$ (due to its (W×L)/10 gate area), thus giving the same $f_c$ we found for the FIRN in the resistor of Fig. 7 in TE. Hence, if the FIRN of the resistor of Fig. 10 in TE was converted into noise voltage "by an $I_{conv}$=150 µA that was able to respect its TE", it would give graph A of excess noise in Fig. 9. But with regard the excess noise of the calibrator of Fig. 10, graph A of Fig. 9 is a kind of "theoretical spectrum" of excess noise in TE that never will appear by setting $I_{conv}$=150 µA in its channel, as it happens with the excess noise of Fig. 6 and the resistor of Fig. 5. In contrast with this, both graphs A and B of Fig. 9 could be obtained with $I_{conv}$=150 µA in the resistor of Fig. 7 by the proper shorting of its gates and its S terminal as we have explained in Section VI.

When $I_{conv}$=150 µA is set in the resistor of Fig. 10, it forces the fingers of its equipotential U-Gate to face regions of the channel whose electrical potentials differ by several units $V_T$. From $V_{DS}$=525mV linearly dropping along x, voltages at positions x=L/5 and x=4L/5 ($v_1$=105mV and $v_2$=420mV respectively) will differ by ($v_2$-$v_1$)=315mV (12.2$V_T$ at room *T*). Thus, the floating U-Gate will acquire some dc voltage $V_{FL}$ accordingly to this situation recalling the gate of Fig. 5 acquiring the $V_{FL}$ of Eq. (17). From the solution of Fig. 8 and our reasoning to obtain its $V_{FL}$ we will say that the p$^+$-n diode under finger 2 of the U-Gate will be reverse-biased, thus extracting a current *i* from the channel. Since this reverse bias will surpass 4$V_T$ (see below) this current will be: *i*≈-$j_{sat}$(W×ΔL). This current leaves finger 2 through the wire to arrive in finger 1 that will inject it in the channel to keep $V_{FL}$ constant. Thus, the diode under finger 1 at x=L/5 will be forward-biased.

Equating injected current at x=L/5 and extracted current at x=4L/5 we have:

$$(W\Delta L)J_{sat}\left[\exp\left(\frac{V_{FL}}{mV_T}\right) - 1\right] \approx (W\Delta L)J_{sat} \Rightarrow V_{FL} = mV_T\ln(2) \quad (21)$$

For *m*=1 and *T*=300K, Eq. (21) gives $V_{FL}$≈18mV≈0,7$V_T$ as the forward bias of the p[+]-n diode of finger 1 at x=L/5. Thus, the reverse bias of the diode under finger 2 is: ($v_2$-$v_1$+$V_{FL}$)=333mV (≈12.9$V_T$) making valid *i*≈*j*$_{sat}$(W×ΔL) for its reverse current that the diode under finger 1 will inject into the channel. This injected current will come from two terms: -*j*$_{sat}$(W×ΔL) that *this diode continues extracting under forward bias* and +2*j*$_{sat}$(W×ΔL) it must inject to have *i*=(2*j*$_{sat}$-*j*$_{sat}$)×(W×ΔL) in this diode. Thus, whereas the shot noise density of the diode under finger 2 drops to $S_{I2}$=2q$I_{sat}$ A²/Hz, that of the diode under finger 1 rises up to $S_{I1}$=6q$I_{sat}$ A²/Hz. Given the low forward bias of this diode ($V_{FL}$=18mV) *its SCR between terminals still is relevant enough* to consider that its two current terms giving rise to its net current *i*=*j*$_{sat}$×(W×ΔL) are *displacement currents* that produce shot noise.

These shot noises under each finger have to do with action **a)** proposed in previous Section to obtain the noise spectrum of each diode from Eq. (20). Thus, whereas the spectrum of Eq. (20) must be halved (-3 dB) for the diode under finger 2, it must be increased by 1.5 (+1,76 dB) for the diode under finger 1. With regard action **b)** the new cut-off frequencies of the noise voltage under each finger are: $f_{c2}$=$f_c$/exp(12.9), thus $f_{c2}$=2,6×10$^{-6}$×$f_c$ for the noise of the diode under finger 2 and $f_{c1}$=$f_c$/exp(-0,7)=2$f_c$ for the noise of the diode under finger 1. Thus, the noise density of the diode under finger 2 will rise by exp(12.9)=3,8×10⁵ (+56 dB increment) as we move its previous $f_c$ to $f_{c2}$=2,6×10$^{-6}$×$f_c$ due to its 3,8×10⁵ times higher resistance, whereas for the diode under finger 1 its noise density will decrease by exp(0,7)=2 (-3 dB reduction) due to its two times lower resistance.

This separated study of each diode is done because the synchronous way they worked in TE to generate noise no longer is possible when $I_{conv}$=150µA is set in Fig. 10. A "synchronous" generation of these Lorentzian noises of $f_{c1}$=0.2Hz and $f_{c2}$=2.6×10$^{-7}$Hz and with their corresponding amplitudes is incredible. From [4, 5], a synchronous generation of noise in the diodes under the fingers of Fig. 10 would require to synchronize their *fluctuations* of electrical energy and this only is possible when the channel is an equipotential region as the floating U-gate is. In this case there is a *common capacity C* between these two equipotential *terminals* for the passage of individual electrons between them as *displacement currents* setting $U$=$q^2$/(2C) in *C*. Because the orthogonal nature of displacement and conduction currents, a "hypothetical fluctuation" mixing a displacement or capacitive current term with a conduction or ohmic one, would not be a fluctuation following [4, 5]. When $I_{conv}$≠0 is set in Fig. 10, the only way to avoid ohmic terms in fluctuations causing noise is to consider separately the capacity of each strip-like diode. This justifies the use of differential, slice-like, JFETs in Fig. 8.

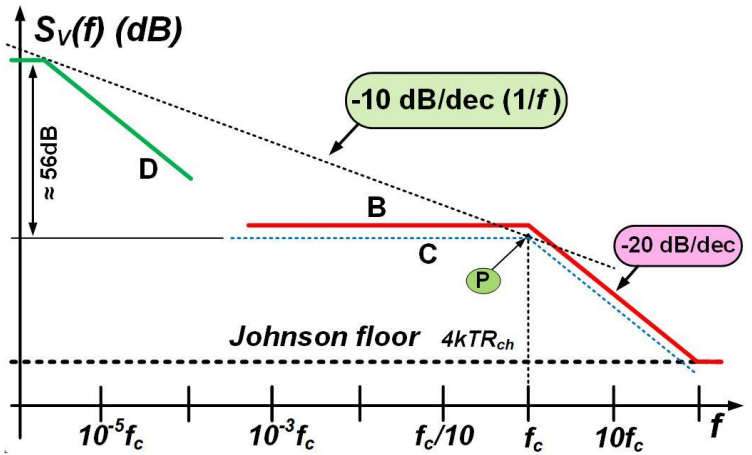

**Figure 11.** Graphical generation of the low-frequency Lorentzian spectrum of excess noise (graph D) that $I_{conv}$=150µA would show in the resistor of Fig. 10.

Taking graph B of Fig. 9 whose link with the FIRN in the short channel of one of the two short FETs of Fig. 7 in TE already was explained, we have done the aforementioned actions **a)** and **b)** to obtain one of the two Lorentzian spectra of *excess noise* appearing between the terminals S and D of the resistor of Fig. 10 when its $I_{conv}$=150µA is set. We refer to the spectrum coming from the noise voltage in the diode under finger 2 that is graph D in Fig. 11. It has been found by lowering 3dB graph B of Fig. 9 accordingly to action **a)**. This gives graph C with dotted lines whose corner in this asymptotic drawing of a Lorentzian (Bode plot) is point P. Next, displacing graph B along the dotted line of -10 dB/decade to put its corner on its cut-off frequency $f_{c2}$=2.6×10$^{-6}$×$f_c$ whose period is T≈44 days, we have completed action **b)**.

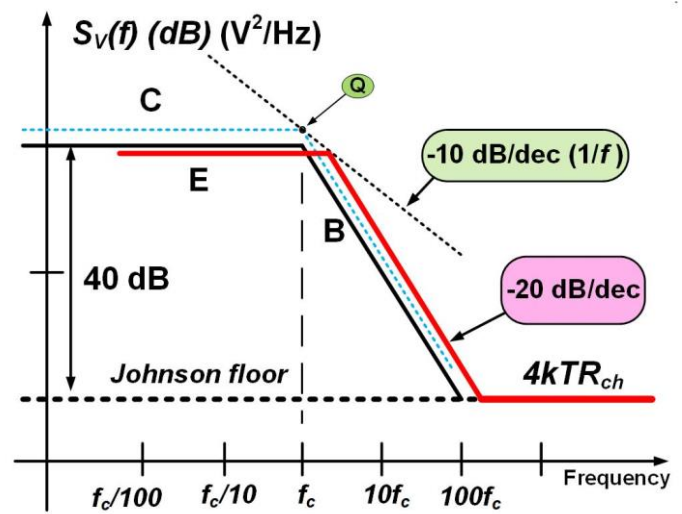

**Figure 12.** Generation of the high-frequency spectrum of excess noise (graph E) that $I_{conv}$=150µA will show in the resistor of Fig. 10 (see the text).

Starting with graph B of Fig. 9 let us carry out the required actions **a)** and **b)** to obtain the second Lorentzian spectrum of excess noise in the calibrator of Fig. 10 under $I_{conv}$=150µA. Fig. 12 shows this spectrum coming from the noise voltage in the diode under finger 1 that requires to shift graph B by +1,76dB to account for its 50% higher shot noise than in TE. This is done in Fig. 12 where graph B shifted by +1.76 dB to accomplish action **a)** is graph C, whose "corner" is point Q. Drawing the dashed line with $1/f$ slope that passes through point Q we have the guide to glide graph C so as to put its corner on the new $f_{c1}=2f_c$ to accomplish action **b)** in this case where the dynamic resistance of the diode is halved by its $V_{FL} \approx 18mV$ forward bias. This gives graph E of Fig. 12.

Fig. 13 shows together the two Lorentzian spectra of excess noise that appear in the resistor of Fig. 10 under its $I_{conv}=150\mu A$. One of these spectra is a hot spectrum of cut-off frequency $f_{c1}=2f_c$, thus not too far from the $f_c$ of the FIRN of this device in TE. Although $f_{c1}=2f_c$ is far from $f_c$, it is "close enough" so as to suggest that it could have to do with our measuring method somehow doubling the $f_c$ of the FIRN that existed in TE without $I_{conv}$. However, when $I_{conv}$ is low this hot spectrum uses to disappear under the Johnson floor. In our case, this is not so because added to the noisy $p^+$-GaAs substrate used, we took a purposely high $I_{conv}=150\mu A$ to show this hot spectrum together with its "cold companion" (graph D) below the µHz well over the Johnson floor in Fig. 13. In contrast with $f_{c1}=2f_c$, the $f_{c2}=0.26\mu Hz$ of the cold Lorentzian lying more than five decades below $f_c$, hardly suggests any link with the Lorentzian FIRN of $f_c=0.1$ Hz that this device had in TE. Without the "guided window" of Fig. 13 nothing would suggest that it is being generated in the same region under finger 2 that generated the 50% of the FIRN of this device in TE whose cut-off frequency was $f_c=0.1$ Hz.

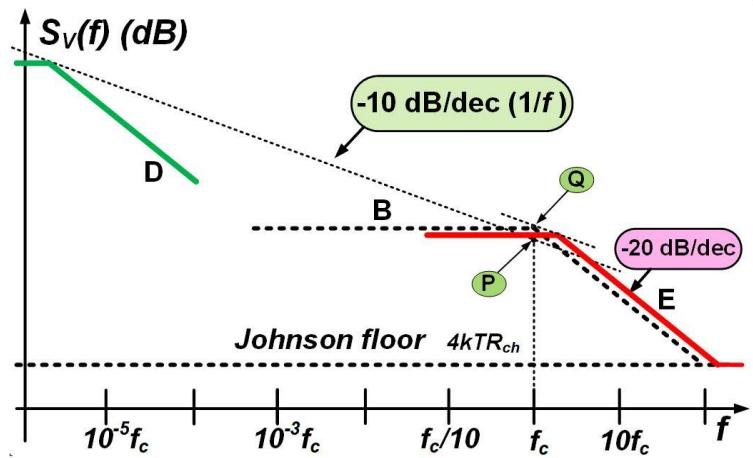

**Figure 13.** Cold (D) and hot (E) Lorentzian spectra of excess noise that $I_{conv}=150\mu A$ would show in the resistor of Fig. 10 from its spectrum of FIRN in TE that would be graph B divided by $(I_{conv})^2$ (see the text).

Previous sentence is particularly true when the hot spectrum lies under the Johnson floor due to a low $I_{conv}$ taken *to reduce the heating* of the resistor *or to reduce its Bias-Induced Departure from TE* (BID) given by the number of thermal units $V_T$ between its terminals. In our case, we have considered that a medium BID≈20 ($V_{DS}=525mV$) is a convenient value for this educational paper despite the nine decades of Lorentzian FIRN spectra ($\exp(BID) \approx 10^9$, see below) that it will give in the resistor of Fig. 5. This care about the BID of devices where excess noise is being measured not always is found in the literature where departures like BID>2000 can be found [17] without a comment on its physical meaning and worse enough: without a word about excess noise that at its time of writing should be well-known [7, 9, 10, 14]. Note the characteristic increase of the noise shown in Fig. 2 of [17] with $(I_{conv})^2$, thus accordingly to Eq. (1).

Once the genesis of the two spectra of Fig. 13 has been shown let us show how its "cold" Lorentzian spectrum of excess noise synthesizes the excess noise with 1/*f* spectrum that everybody finds when a current $I_{conv}$ is injected to convert resistance noise into excess noise. For this purpose, let us add a third finger at x=3L/5 to the resistor of Fig. 10, thus between its previous fingers, but closer to finger 2 than to finger 1. Shorting these fingers as shown in Fig. 14 we will obtain a third spectrum of excess noise in the new "m-Gated" calibrator thus obtained.

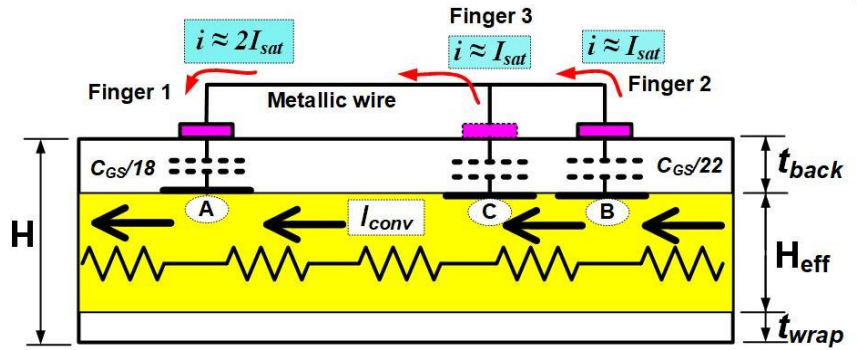

**Figure 14.** Sketch of the third finger added to the resistor of Fig. 10 to get a "warm" spectrum of excess noise added to those of Fig. 13.

From the linear drop of $V_{DS}$=525mV along the channel, voltages at points A, B and C of Fig. 14 would be: $v_1$=105mV at x=L/5, $v_2$=420mV at x=4L/5 and $v_3$=315mV at x=3L/5. The "capacities" $C_{GS}$/18 and $C_{GS}$/22 written in Fig. 13 only are "figures" to recall the small changes of the junction capacity due to the NUB of the diode. Assuming that diodes under fingers 2 and 3 are reverse biased and taking this floating m-Gate as an equipotential region, let us obtain the voltage $V_{FL}$ that it will acquire accordingly to this new situation recalling the $V_{FL}$ of the long gate of Fig. 5 under $I_{conv}$=150µA. Considering that the current that finger 1 injects into the channel is the sum ≈2(L×W)×$j_{sat}$ of the current that diodes under fingers 2 and 3 extract from the channel, Eq. (21) becomes:

$$(W\Delta L)J_{sat}\left[exp\left(\frac{V_{FL}}{mV_T}\right) - 1\right] \approx 2(W\Delta L)J_{sat} \Rightarrow V_{FL} = mV_T\ln(3) \qquad (22)$$

For *m*=1 and *T*=300K, Eq. (22) gives $V_{FL}$≈28mV≈1,1$V_T$ as the forward bias of the p$^+$-n diode of finger 1 at x=L/5. From this $V_{FL}$ the *reverse* bias of the diode of finger 2 will be: $(v_2-v_1+V_{FL})$=343mV≈13.2$V_T$ and the *reverse* bias of the diode of finger 3 will be: $(v_3-v_1+V_{FL})$=238mV≈9.2$V_T$. Since these reverse bias surpass 4$V_T$, our approach $i$≈$j_{sat}$(W×ΔL) for their reverse currents is valid. In this way these two diodes extract a current $i$≈2$j_{sat}$(W×ΔL) that the diode under finger 1 has to inject into the channel as a *net current* $i$≈2$j_{sat}$(W×ΔL). This net current will come from the reverse current -$j_{sat}$(W×ΔL) that remains under $V_{FL}$ (extraction term) and from +3$j_{sat}$(W×ΔL) required to inject $i$=(3$j_{sat}$-$j_{sat}$)×(W×ΔL) in the channel. Thus, the shot noise density of each diode under fingers 2 and 3 will be: $S_{I2}=S_{I3}=2qI_{sat}$ A$^2$/Hz (half their values in TE) whereas the shot noise density of the diode under finger 1 will be twice its value in TE (i. e. $S_{I1}=8qI_{sat}$ A$^2$/Hz).

From these values we observe that the cold spectrum of Fig. 13 becomes "*colder*" due to the new $V_{FL}=28$mV of this m-Gate (10 mV higher than $V_{FL}=18$mV in the U-gate). From this new $V_{FL}$ the cut-off frequency of the cold spectrum is: $f_{c2}=1.82\times10^{-6}f_c$, thus $f_{c2}=0.18\mu$Hz (T≈64 days) and its corner also will be on the same line of -1/*f* slope passing through point P in Fig. 11. With regard the hot spectrum of excess noise due to the forward biased diode of finger 1, its cut-off frequency becomes: $f_{c1}=3f_c$ whereas its amplitude is ≈0.56dB lower than that of graph E in Fig. 13, which was obtained after +1,76dB increment (due to its 150% more shot noise than in TE) and a -3dB decrement due to its 2 times lower dynamical resistance (thus -1.24dB in all). For the new m-Gate with three fingers we have considered that the noise of the diode under finger 1 would be obtained after a +3dB increment due to its 200% more shot noise followed by -4.8dB decrement of due to is 3 times lower dynamical resistance (thus -1.8dB in all). From (1.8-1.24)=0.56dB we justify the new amplitude of graph E in Fig. 15.

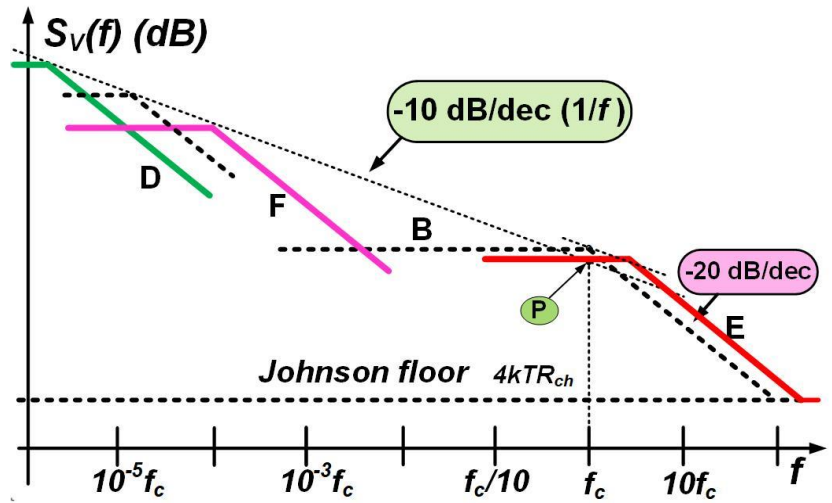

**Figure 15.** Synthesis of 1/*f* excess noise in a resistor by a set of Lorentzian spectra coming from the backgating noise and the instrumental disturbance [6] that occurs when we convert resistance noise into excess noise by a current $I_{conv}$ (see the text).

Concerning the "warm" Lorentzian of excess noise due to finger 3, its cut off frequency would be: $f_{c3}=f_c/\exp(9.2)$, thus $f_{c3}=10^{-4}\times f_c$ and its corner also would be on the dotted line of -1/*f* slope passing through point P of Fig. 11. This is graph F of Fig. 15, where graph D is the updated cold spectrum with $f_{c2}=0.18$ μHz and graph E is the updated hot Lorentzian with $f_{c1}=3f_c$ of the diode under finger 1. A fourth finger midway between fingers 2 and 3 would give the fourth Lorentzian spectrum between D and F spectra (the dashed, non labelled one of Fig.15). Due to this fourth finger, spectrum E of Fig. 15 would have $f_{c1}=4f_c$ and its amplitude would be that of Fig. B decreased by 10log(2.5/4)=-2dB given its 2.5 times higher shot noise and its 4 times higher $f_c$ ($f_{c1}=4f_c$). And so on adding more fingers….

The way these cold Lorentzian spectra of excess noise generated by $I_{conv}$ align to form a "ladder" of 1/*f* slope in Fig. 15 shows the *synthesis of excess noise with 1/f spectrum* that produces in resistors the employment of $I_{conv}$ to convert resistance noise into excess noise. For a "comb-Gate" with *m* narrow fingers connected by (*m*-1) wires to approach the continuous gate of Fig. 5, we can guess the result found by the continuous treatment done in [6] for the excess noise of the resistor of Fig. 5. As a numerical example of this "ladder", Fig. 16 shows three decades of 1/*f* excess noise synthesized by ten Lorentzian spectra aligned in this way and whose $f_c$ values are: 3, 6.5, 13.9, 30, 64.6, 139.2, 300, 646.3, 1392,5 and 3000Hz (dashed lines) [13]. The 1/*f* spectrum they produce (thick line) would run slightly over the dashed line with -10 dB/dec slope of Fig. 15 due to the π/2 factor (≈2dB) due to the summing integral of FIRN along the channel [6]. All the above shows why injecting $I_{conv}$=150µA in the resistor of Fig. 5 will not give the Lorentzian excess noise of Fig. 6, but the 1/*f* excess noise in accordance with Hooge's formula that Fig. 1 shows.

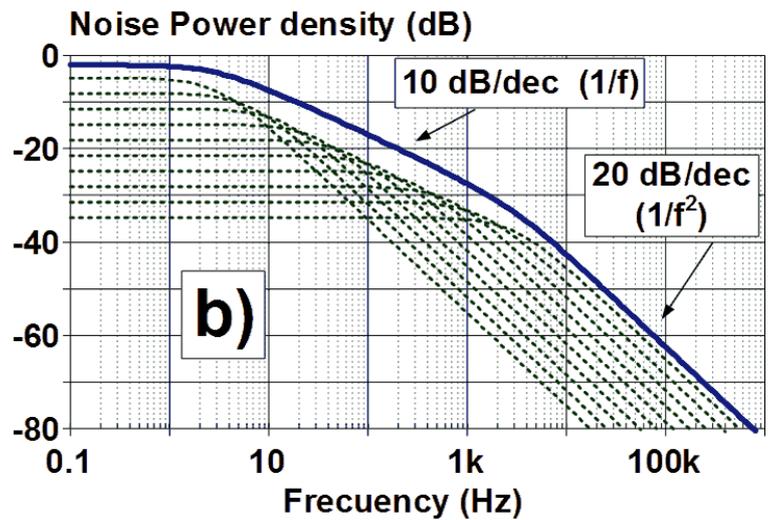

**Figure 16.** Set of ten Lorentzian terms of FIRN synthesizing three decades of 1/*f* FIRN noise [13].

Note that the alignment of Lorentzian terms of FIRN along a 1/*f* line (i. e. their "proper weight" to synthesize a 1/*f* spectrum) naturally occurs because all of them come from similar densities of shot noise (W×ΔL)×2q$I_{sat}$ A$^2$/Hz) driving the admittance of its portion (W×ΔL)/L of BSCR inducing FIRN in the channel. This would be a *characteristic feature of the excess noise in resistors driven by a dc current*, not only in planar ones but also in those containing BSCRs and SCRs affecting their conductive channels. Thinking of the BSCR (double layer) between an insulating support (material 1) and a highly conducting metallic film covering it (material 2) we can foresee a weak 1/*f* excess noise for resistors made from such metallic film, but *not a null excess noise* because FIRN and its associated excess noise is a *proximity effect* induced in the channel of a 2TD from neighbor bodies. This explains the ubiquity of 1/*f* excess noise in electronic devices containing conductive channels like resistors, FETs, BJTs and MOSTs.

It is worth noting the high 1/*f* excess noise that devices using accumulation-channels like MOS transistors should show. In this case, a thin channel that did not exist in TE, is formed and sustained in time by the field-effect induced from a properly biased gate. Let us put this gate over a thin oxide layer grown on the top surface of a semiconductor wafer. As it is well-known this type of channel is a very thin one, thus very prone to show FIRN due to its low $H_{eff}$, see Eq. (11). Although a low-impedance voltage generator biasing its metallic gate could exert a tight control of the voltage in the upper capacity $C_{GS}$ that appears between the gate on top and the thin channel created by this voltage, such generator nothing can do to "silence" the thermal noise that thermal activity sets in the bottom side of the channel. We refer to the $kT/C_{SS}$ noise of the capacitor $C_{SS}$ that is born between the bottom side of the channel and the underlying substrate. Thus, the channel electrically built in the MOST has two "faces": the upper one that would be quite "stable" due to the aforementioned generator and the lower face that would be a "trembling" one as dictated by the thermal activity in $C_{SS}$. All in all, MOSTs would have thin, non-rigid channels, thus bearing a high FIRN from the substrate without needing any "mobility noise" like that considered in [15].

Updating Fig. 2 with the 1/f spectra of $S_R(f)$ deduced from the 1/f excess noise synthesized in Fig. 15 as the number of fingers approaches the continuous gate of the resistor of Fig. 5, one obtains Fig. 17. This $S_R(f)$ represented by the thick dotted line running 2dB over the upper border of Hooge's formula set by $\alpha_H=10^{-3}$ means that the n-GaAs resistor of Fig. 10 having a "comb-like" gate with many fingers to emulate a continuous gate on its channel, would perfectly follow Hooge's formula with $\alpha_H=(\pi/2)\times10^{-3}$, very close to its initial value $\alpha_H=2\times10^{-3}$ [7].

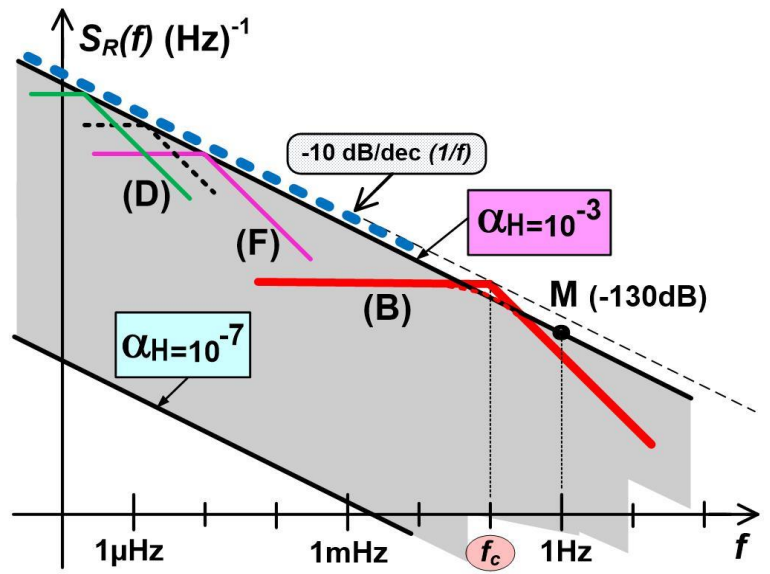

**Figure 17.** Updated version of Fig. 2 showing the normalized density $S_R(f)$ (thick dotted line) that is deduced from the 1/f excess noise synthesized by the Lorentzian terms of Fig. 15. (recall Fig. 2 and its (B) graph).

Since this "comb-gate" simply replaces the p$^+$-GaAs substrate (continuous gate) left floating in the resistor of Fig. 5, the thick dotted line of Fig. 17 is the $S_R(f)$ we would deduce from its excess noise measured under $I_{conv}$=150µA (and also using other $I_{conv}$ values, given the *normalized form* of Hooge's formula). This justifies why the excess noise of the resistor of Fig. 5 for $I_{conv}$=150µA would not show the spectrum of Fig. 6 mirroring its FIRN in TE. This device would not be *in TE*, but under a BID=20 caused by $I_{conv}$=150µA and the way this device express this departure from TE is by the eight decades of excess noise with 1/f spectrum that it would show below the Hz.

## VIII- Working with the new notions and learning from the past

To enter the $S_R(f)$ of our resistor into the shadowed region of Fig. 17, let us obtain the noise reduction we would obtain by replacing its $p^+$-GaAs substrate by a lightly doped one. Using p-GaAs with $N_a=10^{16}$ cm$^{-3}$ acceptors under our n-GaAs layer with $N_d=10^{17}$ cm$^{-3}$ donors, the one-sided BSCR of the p-n$^+$ junction thus formed mostly would lie in the substrate side, and though the built-in voltage of this p-n$^+$ diode would be lower than $\Phi\approx1.3$ V already used, let us keep this value for a rough estimation of $t_{BSCR}$, the thickness of this new BSCR. Replacing $N_d$ by $N_a$ in Eq. (8) we have: $t_{BSCR}=\sqrt{10}\times t_{back}$ and from the ratio $N_d/N_a=10$ the region of this BSCR entering the n-GaAs channel roughly would be ten times lower, thus: $t_n\approx t_{back}/\sqrt{10}$. Due to $t_n<t_{back}$, the previous channel thickness $H_{eff}=0.75\mu m$ would increase up to $H_{eff}\approx0.85\mu m$. From Eq. (11), the first reduction factor $(85/75)^2$ due to this thicker channel would be 1.1dB.

Concerning fluctuations of $t_n$ affecting the channel ($\Delta t_n$) the apparent 20dB attenuation factor expected from $<(\Delta t_n)^2>\approx<(\Delta t_{BSCR})^2>/100$ really would be 10dB lower due to the $\sqrt{10}$ times larger thickness of the new BSCR giving rise to $\sqrt{10}$ times larger fluctuations ($\Delta t_{BSCR}=\sqrt{10}\times t_{back}$) for similar voltage changes. Since the relative changes $\Delta t_{BSCR}/t_{BSCR}$ and $\Delta V/(2\Phi)$ are equal, see Eq. (10), this means that a voltage fluctuation $\Delta V$ giving rise to a fluctuation $\Delta t_{back}$ in the channel when the substrate was $p^+$-GaAs would produce $\Delta t_n\approx\Delta t_{BSCR}/10$, thus $\approx\Delta t_{back}/\sqrt{10}$ in the n-GaAs channel onto this lightly doped substrate. In this way the noise reduction would be 10dB at first sight for the same amount of $kT/C_{GS}$ noise creating FIRN. Unfortunately, the achieved reduction would be lower because $t_{BSCR}=\sqrt{10}\times t_{back}$ means $\sqrt{10}$ times *lower capacity* between channel and substrate, thus $\sqrt{10}$ higher $kT/C$ noise in Eq. (11). All in all, a noise reduction of only 5dB results that, added to the 1,1dB reduction of previous paragraph, gives an overall reduction close to 6dB (thus four times lower power).

The "better technology" we obtain by using lightly doped p-GaAs substrates would give a 3.5k$\Omega$ resistor with four times lower excess noise power. Looking at the corner frequency $f_{CN}$ where its 1/$f$ excess noise would cross its Johnson floor, we would have reduced by 4 its previous value. This notion of bad technology linked with higher excess noise in devices is a familiar one that has led to use excess noise as a tool for the quality assessment of materials [18]. The 6dB less excess noise power using the lightly doped substrate would enter the $S_R(f)$ of our resistor well into the grey band of Fig. 17 and moreover, the noise estimation just done could be extended to devices onto SI-GaAs substrates like those of [11] or like those used in MMICs (Microwave Monolithic Integrated Circuits).

From the Instrumentation viewpoint the main lesson to be learned would be: "*everybody finds 1/f excess noise accordingly to Hooge's formula because everybody uses a current to measure resistance noise relying on Eq. (1)*" as we did in 1998 for resistors made from InGaAs/GaAs buffer layers [19]. At that time we had not an idea on the origin of the 1/$f$ excess noise, but in subsequent years we never forgot the *excellent generators of 1/f excess noise* that these highly *cross-hatched* "samples" were, as Fig. 2 of [19] shows. Due to our preliminary acquaintance with resistance noise at that time we could not give then a good reason for this striking feature. From the widespread use of the word "sample" replacing "device" in [19] and the title of its Section II: "Sample description" we recognize our prejudice of excess noise coming from "bulk regions of materials" likely guided by works like [7] that we took as classical ones we had to follow.

Leaving aside this prejudice was a progressive task that gave us a good reason for the high generation of 1/f excess noise in our devices that is the high amount of FIRN caused by the high number of BSCRs "coherently working" (that is: *spatially aligned*) along the channels of these cross-hatched resistors we were measuring. This reason agrees with the high excess noise of devices that use granular or polycrystalline materials and with the higher photo-conductance that we proposed for a set of δ-doped layers embedded in photoconductors [11] to have sheets of positive donor ions along the channel screened by "sheet-like" clouds of free electrons as conductive channels between terminals. However, we are afraid that our model for 1/f *excess noise* that we distinguish from *flicker noise* (flux noise with spectrum close to 1/f studied in [8]), is quite apart from the current opinion on the "1/f problem" as we wrote at the end of the Introduction, thus being hard to accept at first sight.

Although most people accept that the 1/f excess noise of a resistor reveals fluctuations $\Delta R$ of its resistance around its mean value $R$ and we fully agree with this idea, the conflict appears because most people believe that the *deduced* $\Delta R$ noise with its particular 1/f spectrum is something coming from the material bulk within the resistor, whereas following our model it only is the channel modulation due to the unavoidable capacity between two neighbor conductors. This capacity meaning that the electric field stores energy between them (Maxwell equations) should be taken into account as well as the degree of freedom that it represents from the thermal viewpoint [10]. Thus, electromagnetism and thermodynamics would give together *the reason* for the ubiquitous 1/f excess noise of resistors: a *backgating noise* with Lorentzian spectrum in general that, in resistors biased by a dc current $I_{conv}$, adopts a characteristic "costume": its *1/f spectrum*.

This solution to the "1/f problem" that would end the endless dispute on mobility or carrier number fluctuations giving rise to 1/f excess noise [9, 14, 15] has to do with a lesson learned twenty years ago [16]: that *what we measure* in a photoconductor is the *conductance* between its terminals, not the *conductivity* of its inner material *that must be inferred* from the former, usually from the rigid channel approach that being useful for static estimations of mean conductivity, becomes useless for tiny fluctuations of resistance. Given the amount of works on the aforementioned dispute, the review of our paper removing its basement could be a non-easy task, but considering that such dispute relies on conductivity that nobody has measured yet, but only inferred from *conductance* measured in two terminal devices, we decided to write this paper to end the aforementioned dispute and to show a theoretical basis under Hooge's empirical formula.

The removal of our prejudice of 1/*f* excess noise coming from bulk regions of material led us to write [6] and this paper, but we are afraid that electrical noise is a field where other prejudices still remain. One of them would be the notion we have on $S_I(f)=4kT/R$ A$^2$/Hz, the spectral density of *noise current* in resistors called their Nyquist noise. This density uses to be associated with small and random currents due to charges agitating between the terminals of a resistor (see the titles of [1, 2]), thus with random currents that an electronic i-v converter *could extract from the resistor* to convert them into a proportional noise voltage on its feedback network driven by its output port. The main drawback of this "intuitive" model of electrical noise is that it is hard to fit into the Fluctuation-Dissipation framework derived from [20]. By contrast, the fitting of [4, 5] in this framework is easy if one keeps the electronic charge *q* as the quantum of charge involved in each fluctuation generating electrical noise in resistors (recall *U=q$^2$/(2C)*, the mean energy of these fluctuations in our model).

Using this quantum model in resistors we find that their Johnson noise (i. e. the *voltage* we find between their two terminals) becomes "the measurable *effect*" of a random set of tiny fluctuations of electric field between terminals, each being equivalent to the displacement of a single electron between them. Since the first news that the feedback electronics has on the occurrence of one of these events (or *cause* of noise) is a voltage step of *q/C* volts on *C* (its *effect* looking like the sudden jump of an electron between terminals), the picture of these displacement currents leaving the resistor through the feedback network of the i-v converter without having created a voltage in *C* driving the feedback electronics governing such network is unbelievable. For impulsive noise like this one, the capacity *C* of a resistor would not allow the current-to-voltage conversion that we would expect to occur in the feedback network of a i-v converter because such conversion would be done "within *C*" although the output signal of the converter will suggest that it took place in its feedback network [3].

Thus, the idea of i-v converters converting "noise current" (i. e. Nyquist or shot noise) into noise voltage would fail for this "quantum" or "impulsive" noise. In fact, anyone having designed this type of converters under the continuous, dissipation-based, noise model for electrical noise, knows that for those noise components of high enough frequency, the capacity *C* between terminals of the resistor becomes a faster i-v converter than the i-v one is designing to connect to the resistor. Thus, why not consider that this happens at any frequency to keep the integrity of *q* in electrical noise processes? [3]. In other words: *what if this impulsive noise model was true*? In this case, the output signal delivered by an i-v converter trying to extract the "noise currents" of a resistor (i. e. trying to convert its Nyquist noise into noise voltage) would be the same output signal it would deliver if it actually was extracting such currents and converting them into voltage, but in fact it only would be *amplifying the tiny voltage noise in C* (*effect*) due to its *causes* "already gone" that would be the meaning of the theoretical Nyquist noise *inferred* from the Johnson noise voltage we measure on *C* [3-5].

If the proposed impulsive noise was true, equipment designed to measure *noise current* in resistors like their shot or Nyquist noise ($A^2/Hz$), really would measure their *noise voltage* on *C*, thus showing their Johnson noise together with any other noise voltage (like excess noise) that a proper stimulus could produce between the terminals of the resistor. Taking Fig. 2 of [17] and Fig. 2 of [19] both graphs show noise *voltage* with 1/*f* spectrum that evolves accordingly to Eq. (1) in two very different resistors. This suggests that both figures are showing excess noise proportional to $(I_{conv})^2$, for $I_{conv}$ being the dc current set in each resistor. However, whereas Fig. 2 of [19] shows noise voltage (the familiar excess noise) Fig. 2 of [17] is said to show noise current (shot noise).

Looking for other relevant differences we can see that whereas Eq. (1) of [19] is Hooge's formula, no mention to Hooge's work is done in [17] despite the striking 1/*f* noise of its Fig. 2 and its characteristic evolution following Eq. (1). Considering the journals where these works appeared, one would be a journal of Electrical Engineering [19] and [17] would be in a journal of Physics. Since a journal of Physics unaware of excess noise sounds strange and irritating, a likely reason for this oblivion could be an "urgent whish" of its authors and referees to show the first empirical evidence of noise that could have to do with the elusive noise density $2qI_{conv}$ $A^2/Hz$ that most people associate today to a conduction current like $I_{conv}$ (as we ourselves would have done few years ago).

Knowing that following [3-5] the first report about shot noise in macroscopic resistors in TE was that of Johnson ninety years ago [1], let us consider with care the noise of [17]. Because the "*noise current* with 1/*f* spectrum" of Fig. 2 of [17] is said to be noise current given in $A^2/Hz$ (Nyquist noise for $I_{conv}=0$) and its power density evolves proportionally to $(I_{conv})^2$ following this figure, these authors are proposing a new shot noise whose density (in $A^2/Hz$) would be proportional to $(I_{conv})^2$. This proposal would depart from the familiar shot noise density $2qI_{conv}$ $A^2/Hz$ [21] that is proportional to the current $I_{conv}$, not to its square. Therefore, the name "shot noise" these authors give to the noise they propose in Fig. 2 is quite misleading and its genesis from the "piling-up" of electrons "nowhere" because *no capacity C was considered* by these authors, is doubtful.

By contrast, we explain the noise of Fig. 2 of [17] by considering first the high capacity *C* of its CdTe resistor coming from the high dielectric relaxation time $\tau_d$ of its inner material [6]. For impulsive noise, electrical noise already would be under the form of a small, fluctuating voltage in this capacity. In this way the noise shown in Fig. 2 would not be noise current before being converted into voltage noise, but *noise voltage appearing in C* that the i-v converters of [17] would be amplifying with a gain suggesting that they actually are converting noise current into noise voltage as shown in [3]. This would justify the observed features in Fig. 2 of [17] like the 1/*f* spectra of these measurements for $I_{conv} \neq 0$ evolving accordingly to Eq. (1) because these i-v converters would be amplifying the Johnson noise born in *C* as well as the excess noise due to the dc current in this resistor. The presence of this high *C* would account for the lack of flatness of the "Nyquist noise" shown for $I_{conv}$=0 that exists at *f*=10kHz and that becomes dramatic between 10kHz and 100 kHz, where the effects of *C* on the feedback networks of their i-v converters cannot be hidden. All in all, Fig. 2 of [17] could be taken as a proof about the impulsive nature of electrical noise in resistors.

Finally, let us say that the FIRN we are proposing has no problem with its integral at low frequencies because its 1/*f* spectrum always becomes flat at a low frequency $f_L$ largely determined by the BID from TE used in the measurements. It is worth noting that for BSCRs in silicon devices, $f_L$ values under the µHz should be found as we could deduce from [22], where excess noise of ten different input stages of operational amplifiers (OA) was measured, giving an estimation for their 1/*f* spectrum down to $10^{-6.3}$ Hz (0.5 µHz). This result is not far from the estimation we can do from the $f_c$≈86kHz we measured in broad area silicon Schottky diodes [12, 13] whose built-in potential would be Φ≈0.6-0.7 V lower than those of the Si p-n junctions that would be generating 1/f excess noise in those input stages. From $f_{cOA1}$=86kHz/exp(0.6V/25.9meV)≈7µHz and $f_{cOA2}$=0.16 µHz for Φ≈0.7 V at room *T*, the results of [22] and our estimations would be cogent, particularly if the temperature that was kept within 0.0001 ºF in [22] was lower than room *T* (sorry we could not read this *T* in [22]). This low, but non null $f_L$ determined by the $f_c$ of the Lorentzian *kT/C* noise generating FIRN in the device, by its BID and by the temperature of the experiment, is an interesting feature of our model that avoids the problems of divergent 1/*f* noises as we advanced in [6].

## Conclusion

As we have shown, *the "1/f noise problem" was a Measurement one, thus of fundamental nature and the circuit approach not only suited, but was required for its solution.*